\documentclass[aps, twocolumn,superscriptaddress]{revtex4-1}
\usepackage[utf8]{inputenc}

\usepackage{lipsum}
\usepackage{braket}
\usepackage{amsmath}
\usepackage{amssymb}
\usepackage{hyperref}
\usepackage{siunitx}
\usepackage[normalem]{ulem}
\usepackage{graphicx}
\usepackage{colortbl}
\usepackage{xcolor}
\newcommand{\rebuttal}[1]{#1}

\begin{document}
\title{Fast, high precision dynamics in quantum optimal control theory}
\author{Mogens Dalgaard}
\affiliation{Department of Physics and Astronomy, Aarhus University, Ny Munkegade 120, 8000 Arhus C, Denmark}
\date{July 2019}
\author{Felix Motzoi}
%\email{f.motzoi@fz-juelich.de}
\affiliation{Forschungszentrum J\"ulich, Institute of Quantum Control (PGI-8), D-52425 J\"ulich, Germany}

\date{July 2019}

\date{\today}

\begin{abstract}
Quantum optimal control theory is becoming increasingly crucial as quantum devices become more precise, but the need to quickly optimize these systems classically remains a significant bottleneck in their operation. Here we present a new theoretical quantum control framework for much faster optimization than the state of the art by replacing standard time propagation with a product of short-time propagators, each calculated using the Magnus expansion. The derived formulas for exact series terms and their gradients, based on earlier approximate integrals in a simulation setting, allow us to subsume the high cost of calculating commutators and integrals as an initial overhead. This provides an order of magnitude speedup \rebuttal{for quantum control optimization}. 
\end{abstract}
\maketitle
\section{Introduction}

Quantum technologies may solve societally relevant problems within areas such as optimization \cite{farhi2014quantum}, equation solving \cite{harrow2009quantum}, drug design \cite{peruzzo2014variational, mcclean2016theory}, and machine learning \cite{schuld2015introduction,biamonte2017quantum} in combinatorially difficult regimes where classical computers are expected to fail \cite{dalzell2020many}. In recent years, we have witnessed the small-scale implementation of some of these ideas \cite{boixo2016computational,kandala2017hardware, kokail2019self,omran2019generation, harrigan2021quantum}. However, large-scale implementations still require dramatic improvement in our ability to accurately simulate and control subparts of these systems. 

Quantum control optimization is formally treated within quantum control theory \cite{d2007introduction,werschnik2007quantum,glaser2015training, Georzthesis}, where recent progress includes: frequency domain optimization \cite{dong2020learning, sorensen2020optimization, muller2021one}, Hessian-based optimization \cite{goodwin2016modified,dalgaard2020hessian,jensen2021approximate}, optimization of many-body matrix product states \cite{jensen2020achieving, omran2019generation}, noise-resilient control \cite{muller2018noise,liu2019plug, khani2012high, gupta2020adaptive, dalgaard2021dynamical}, reinforcement learning-based optimization \cite{palittapongarnpim2017learning, bukov2018reinforcement,xu2019generalizable,an2019deep,dalgaard2020global, yao2020policy, baum2021experimental}, circuit optimization \cite{mitarai2018quantum, motzoi2017linear, arrazola2019machine}, feedback control \cite{chen2020quantum, magrini2021real, motzoi2016backaction, basilewitsch2019reservoir}, and global cost functional landscape optimization \cite{kokail2019self, koczor2020quantum, dalgaard2021predicting}.

\rebuttal{
In the context of quantum simulations, a widely used approach is the rich class of numerical differential equation solvers such as Runge-Kutta \cite{butcher1996history}. However, state-of-the-art quantum control optimization relies on gradient-based minimization, which often outperforms gradient-free alternatives \cite{sorensen2018quantum}. Although calculating the gradient with, e.g., Runge-Kutta is possible \cite{machnes2018tunable}, it is usually a numerically expensive and slow approach. For this reason, we explore in this paper gradient-based optimization that relies on the Magnus expansion \cite{magnus1954exponential, blanes2009magnus, auer2018magnus, ma2020optimal}, which benefits from significantly better convergence properties than, e.g., Taylor or Dyson series.} However, like these other series, going beyond a few orders is computationally (or analytically) taxing, while lower-order results only provide approximate solutions. \rebuttal{Therefore a standard approach is to resolve the total simulation duration into shorter, high-accuracy time intervals, which naturally leads to a trade-off between accuracy per time step and the total number of time steps. }

\rebuttal{Within quantum control optimization, a current state-of-the-art gradient-based algorithm, GRAPE \cite{khaneja2005optimal,de2011second}, is a low-accuracy (per time step) method, viewable as an approximation of the first term in the Magnus expansion. In this work, we demonstrate that one may easily improve GRAPE by including higher-order terms in the Magnus expansion. 
Our contribution is thus two-fold: (i) we derive the control gradient for three different integrations schemes based on the Magnus expansion. (ii) We compare the performance of these four (i.e., including standard GRAPE) control optimization methods in equal computational resource comparisons on both slow and fast dynamics, relative to the control fields' modulation rate.}

\rebuttal{
A crucial insight of our work is the ability to pre-calculate commutators and integrals, allowing for the fast inclusion of higher-order terms in the Magnus expansion. Our method also leads to the option of whether to approximate or exact calculate the arising integrals. The latter approach increases accuracy per time step but at the cost of overhead in either human labor or computational resources. This overhead is seldom justified for standard ``single-shot'' simulations, making Magnus-based integration rarely more efficient than other frequently used integration techniques. However, quantum control optimization generally relies on many time propagations, making this overhead overall insignificant.  }

As a testbed for our methods, we optimize the global control of an Ising-type spin chain, recently studied also in Refs.~\cite{dalgaard2021predicting,dalgaard2021dynamical}. Locally interacting spin systems are compelling to study because of their importance within solid-state physics \cite{plischke1994equilibrium}. In addition, many quantum computational architectures depend on adjacently coupled qubits \cite{cirac1995quantum, bloch2008quantum, blais2020quantum,wu2020concise, kinos2021roadmap}, while also having potential applications in surface error correction codes \cite{fowler2012surface,barends2014superconducting} and transport of quantum information \cite{ChristandlStateTransfer,yung2006quantum}. \rebuttal{Our results demonstrate up to a factor of ten in differences in wall time consumption between the best- and worst-performing methods in equal-accuracy comparisons on both fast and slow dynamics.}

The paper is organized as follows: In Section~\ref{sec:theory} we demonstrate how to use the Magnus expansion to derive different integration schemes and how to utilize these for quantum control optimization. In Section~\ref{sec:results}, we benchmark these against each other with respect to both speed and accuracy. In Section~\ref{sec:conclusion}, we present concluding remarks.

\section{Theory} \label{sec:theory}

\begin{figure*}
    \centering
    \includegraphics{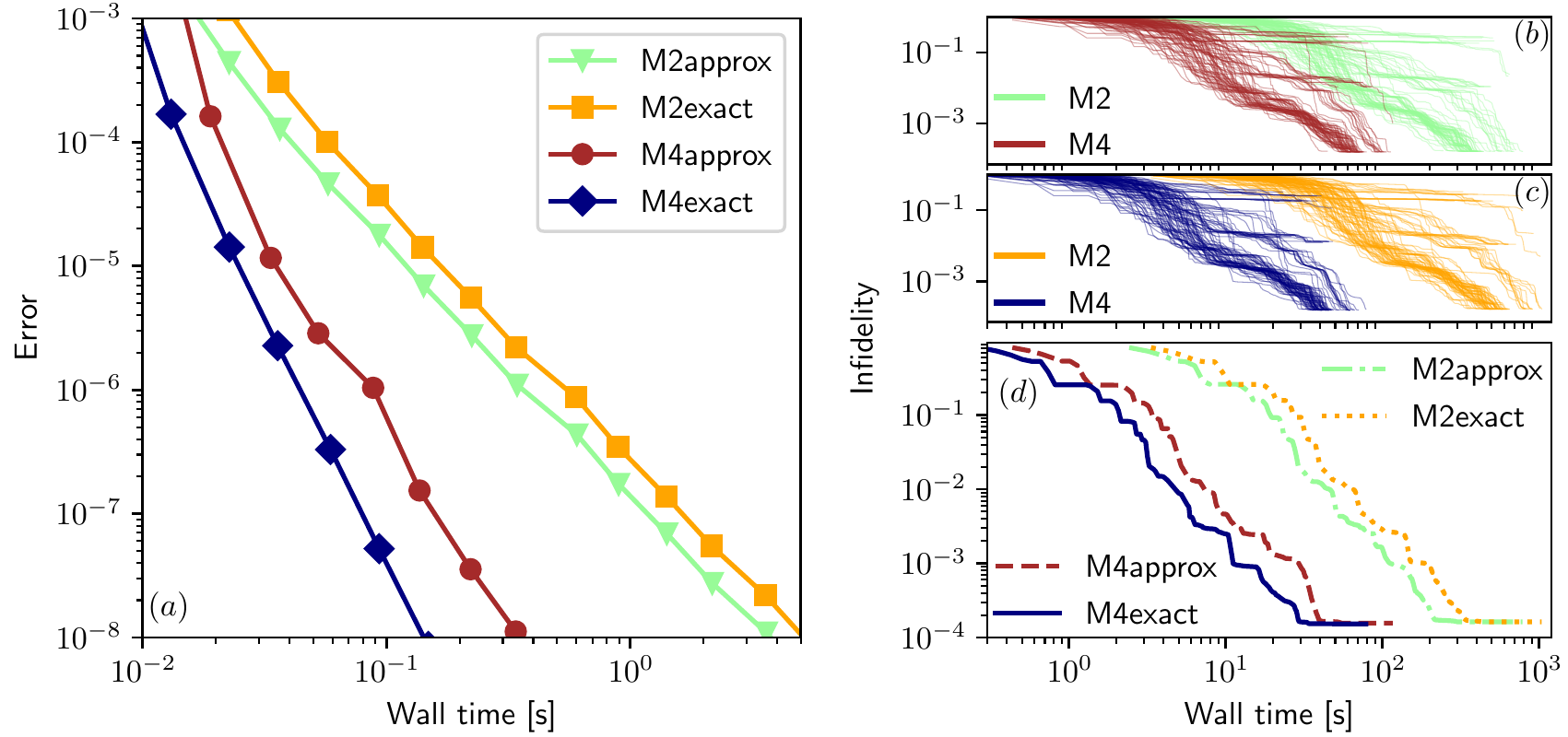}
    \caption{Optimization results with the rotating wave approximation. (a) The wall time consumption for calculating the infidelity versus the numerical error of the infidelity averaged over 100 optimized pulses for the different methods presented in the text. We obtained the data by scanning over different values of $\Delta t$. From this data, we find the $\Delta t$ that leads to an average error of $10^{-6}$. In (b), we compare the optimization from the found discretization of the two approximation methods. We repeat in (c) the comparison for the two exact methods. Finally, in (d), we depict the smallest infidelity found at each wall time duration for each of the different methods.}
    \label{fig:magnus_spin_chain}
\end{figure*}

Quantum control optimization allows for shaping externally applied control fields in order to realize dynamical operations such as state-to-state transfers, unitary gate synthesis, or preparation of a density matrix in both closed and open quantum systems \cite{machnes2011comparing}. One of the most successful approaches in this regard, is the gradient pulse engineering algorithm (GRAPE) \cite{khaneja2005optimal,de2011second} and its many extensions \cite{motzoi2011optimal,schulte2011optimal, goodwin2016modified, Lucarelli2018, sorensen2018quantum, dalgaard2020hessian}. The general idea is to reformulate the control problem as a minimization task of a cost-function $\mathcal{F}[u^{(k)}]$ with respect to a set of control fields $\{u^{(k)} \}_k$ that steers the system via a control Hamiltonian
\begin{align}
    H(t) = H_0 + \sum_k u^{(k)}(t) H_k.
    \label{eq:control_Hamiltonian}
\end{align}
Here $H_0$ and $H_k$ denote drift and control Hamiltonians, respectively, which we assume are time-independent. In this work, we specifically seek to optimize pulses expressed in a set of basis functions $u^{k}(t) = \sum_n b_n^{(k)}\phi_n^{(k)}(t)$. This type of parametrization has been popularized in literature through the chopped random basis approach \cite{caneva2011chopped,muller2021one,van2016optimal,motzoi2011optimal,sorensen2018quantum,sorensen2020optimization}, which is suitable to obtain pulses that are easy to implement experimentally with vanishing effects \cite{sorensen2020optimization} of filtering. With this parametrization, we now seek to minimize the cost-function $\mathcal{F}(\{ b_n^{(k)} \})$ over the pulse parameters $\{ b_n^{(k)} \}$.
\rebuttal{
For gradient-based optimization we also need the derivative $\partial_{b_n^{(k)}} \mathcal{F}$, which for a state-to-state transfer depends on the end state and $\ket{\psi(T)}$ its derivative $\partial_{b_n^{(k)}} \ket{\psi(T)}$ \cite{khaneja2005optimal}, where $T$ denotes the total control duration. By using the Schrödinger equation and its derivative we obtain the coupled equations $(\hbar = 1)$ \cite{Georzthesis, machnes2018tunable}
\begin{align}
    \label{eq:Schrodinger_eq}
    \partial_t \ket{\psi(t)} &= -i H(t)\ket{\psi(t)}\\
    \label{eq:schrodinger_deriv}
    \partial_t \partial_{b_n^{(k)}} \ket{\psi(t)} &= -i \phi_n^{(k)}(t) H_k\ket{\psi(t)}
    -iH(t) \partial_{b_n^{(k)}} \ket{\psi(t)},
\end{align}
which we may solve with, e.g., Runge-Kutta. Unfortunately, this requires propagating an additional state $\partial_{b_n^{(k)}} \ket{\psi(t)}$ per pulse parameter $b_n^{(k)}$, which is much more costly than the usual alternative \cite{khaneja2005optimal}, thereby making Runge-Kutta or similar techniques less tractable for quantum control optimization. In addition, Runge-Kutta does not preserve normalization, which may cause a nonphysical loss of state norm, which is only negligible at a sufficiently high number of time steps. }

Instead we may numerically solve, e.g., the Schrödinger equation, Eq.~(\ref{eq:Schrodinger_eq}), for a duration $T$ in a series of $N$ equidistant steps, $\Delta t = T/N$, via the time evolution operator $U(\Delta t, t) \ket{\psi(t)} = \ket{\psi(t + \Delta t)}$. The time evolution operator for the $j$th time step $t_j = j\Delta t$ is formally given by $U(\Delta t, t_{j-1}) = U_j = e^{-i\Omega_j}$, where $\Omega_j = \sum_{n=1}^\infty \Omega_j^{[n]}$ denotes the Magnus expansion \cite{magnus1954exponential, blanes2009magnus}, where the first term reads
\begin{align} 
    \Omega_j^{[1]} = \int_{t_j}^{t_j + \Delta t} dt_1 H(t_1)
    = \Delta t H_0
    + \sum_k c^{(1)}_{k,j} H_k.
    \label{eq:magnus_first}
\end{align}
Note in the above equation we have inserted Eq.~(\ref{eq:control_Hamiltonian}) and implicitly carried out the integral
\begin{align}
    c^{(1)}_{k,j} = \int_{t_j}^{t_j + \Delta t} dt_1
    u^{(k)}(t_1)
    \label{eq:first_integral}
\end{align}
The second term in the Magnus expansion reads
\begin{align} \nonumber
    \Omega_j^{[2]}
    &=
    - \frac{i}{2}
    \int_{t_j}^{t_j + \Delta t} dt_1 \int_{t_j}^{t_1} dt_2
    [H(t_2), H(t_1)]\\
    &=
    -i\sum_k c^{(2)}_{k,j} [H_0, H_k]
    -i \sum_{k<k'} c^{(3)}_{k,k',j} [H_k, H_k'],
    \label{eq:magnus_second}
\end{align}
where we similarly to before have collected the integrals
\begin{align}
    c^{(2)}_{k,j} =& \frac{1}{2}
    \int_{t_j}^{t_j + \Delta t} dt_1
    \int_{t_j}^{t_1} dt_2
    \big(
    u^{(k)}(t_1)-u^{(k)}(t_2)
    \big)
    \label{eq:second_integral}
    \\ \nonumber
    c^{(3)}_{k,k',j} =& 
    \frac{1}{2}
    \int_{t_j}^{t_j + \Delta t} dt_1
    \int_{t_j}^{t_1} dt_2
    \bigg(
    u^{(k)}(t_2)u^{(k')}(t_1)\\
    &-
    u^{(k')}(t_2)u^{(k)}(t_1)
    \bigg).
    \label{eq:third_integral}
\end{align}
Note, that using the parametrization $u^{k}(t) = \sum_n b_n^{(k)}\phi_n^{(k)}(t)$, we may precalculate the various integrals over the basis functions $\phi_n^{(k)}$, which allows for significantly faster computation. We elaborate further on this in Appendix~\ref{sec:different_methods}.
Each increasing term in the Magnus expansion grows in complexity, hence, 
we must for all practical purposes truncate the Magnus expansion at some finite term $m$. Here the per step error $||\Omega_j - \sum_{n=1}^m \Omega_j^{[n]} ||$ scales as $\mathcal{O}(\Delta t^{2m+1})$ and the accumulated error in the final state $\ket{\psi(T)}$ scales as $\mathcal{O}(\Delta t^{2m})$ \cite{blanes2009magnus}. Note that written in the form of Eq.~(\ref{eq:magnus_first}) and (\ref{eq:magnus_second}), we may also pre-calculate the commutators, which again allows for faster computation of the evolution.

We may obtain the derivatives by using the multi-variate chain-rule similarly to Ref.~\cite{motzoi2011optimal, sorensen2018quantum},
\begin{align} \nonumber
    &\frac{\partial \mathcal{F}}{\partial b_n^{(k)}}
    =
    \sum_j
    \Bigg(
    \frac{\partial \mathcal{F}}{\partial c_{k,j}^{(1)}}
    \frac{\partial c_{k,j}^{(1)} }{\partial b_n^{(k)}}
    +
    \frac{\partial \mathcal{F}}{\partial c_{k,j}^{(2)}}
    \frac{\partial c_{k,j}^{(2)} }{\partial b_n^{(k)}}\\
    &+  
    \sum_{k'<k}
    \frac{\partial \mathcal{F}}{\partial c_{k',k,j}^{(3)}}
    \frac{\partial c_{k',k,j}^{(3)} }{\partial b_n^{(k)}}
    +
    \sum_{k'>k}
    \frac{\partial \mathcal{F}}{\partial c_{k,k',j}^{(3)}}
    \frac{\partial c_{k,k',j}^{(3)} }{\partial b_n^{(k)}}
    + \ldots
    \Bigg)
    .
\end{align}
Here the derivatives $\partial \mathcal{F}/ \partial c_{k,j}^{(1)}$, $\partial \mathcal{F}/ \partial c_{k,j}^{(2)}, \ldots$ are calculable using standard GRAPE propagation  \cite{khaneja2005optimal,de2011second}, \rebuttal{which requires two propagations: one forward and one backwards. This is significantly smaller than the alternative approach, Eq.~(\ref{eq:Schrodinger_eq})  and (\ref{eq:schrodinger_deriv}), which requires one additional propagation of $\partial_{b_n^{(k)}} \ket{\psi(t)}$ per pulse parameter $b_n^{(k)}$ thereby making the alternative much more expensive.}

In this work, we benchmark four different integration schemes for quantum control: two-second order methods with one explicitly calculating $\Omega_j^{[1]}$ (M2exact) and one approximating it via midpoint interpolation (M2approx). As well, we study two fourth-order methods with one explicitly calculating $\Omega_j^{[1]}+\Omega_j^{[2]}$ (M4exact) and one approximating it via Gauss–Legendre quadrature (M4approx). Note, that M2approx and M4approx have previously been used as numerical integration schemes \cite{blanes2009magnus}, while M2approx may readily be seen as the current standard (GRAPE) in the quantum control literature \cite{khaneja2005optimal,de2011second,machnes2011comparing,sorensen2018quantum,dalgaard2020hessian}. \rebuttal{Besides M2approx, none of these methods have, to the best of our knowledge, been used to perform gradient-based quantum control optimization before. As mentioned in the introduction, this paper aims to derive the control gradient needed for state-of-the-art optimization for these three alternatives to standard GRAPE and benchmark all of these methods against each other for control optimization. We elaborate on the more technical details of the former in Appendix~\ref{sec:different_methods} and provide the comparison in the following section. }

\begin{figure*}
    \centering
    \includegraphics{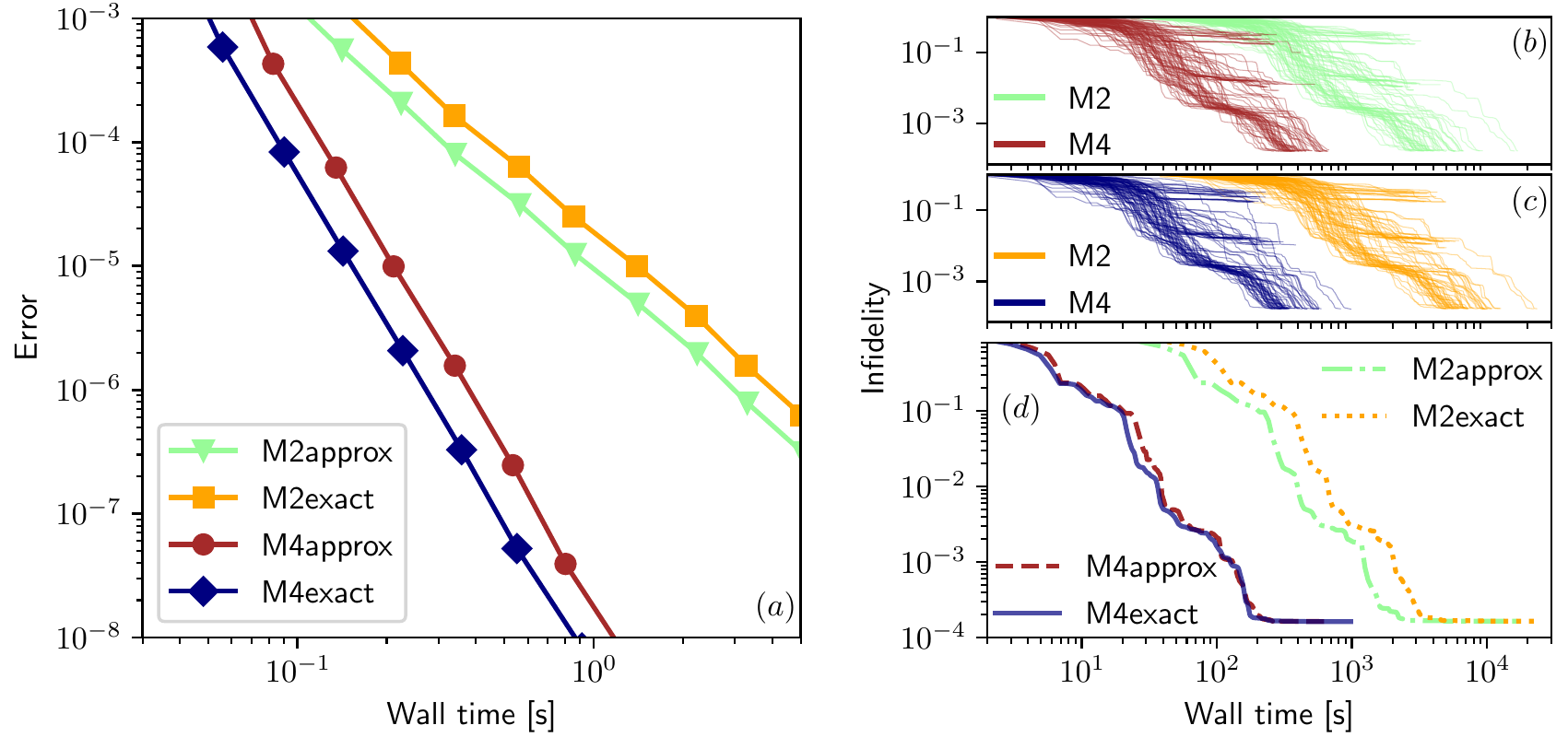}
    \caption{Optimization results without the rotating wave approximation. (a) The wall time consumption for calculating the infidelity for a state transfer on a spin chain without the rotating wave approximation plotted against the numerical error of the infidelity averaged over 100 optimized pulses for the different methods presented in the text. We obtained the data by scanning over different values of $\Delta t$. From this data, we find the $\Delta t$ that leads to an average error of $10^{-6}$. In (b), we compare the optimization from the found discretization of the two approximation methods. We repeat in (c) the comparison for the two exact methods. In (d), we depict the smallest infidelity at each wall time duration for each of the different methods.}
    \label{fig:magnus_spin_chain_no_rwa}
\end{figure*}

\section{Results} \label{sec:results}

\subsection{Controlling a spin chain within the RWA}

To compare the different methods, we consider control of a one-dimensional spin chain where we model an Ising-type spin-spin interaction for nearest and next-nearest neighboring spins, which was also studied recently in the context of optimal control theory in Refs.~\cite{dalgaard2021predicting,dalgaard2021dynamical}. We assume access to two independent global control fields $u^{(x)}$ and $u^{(y)}$ orthogonal to the chain and driven at qubit-resonance. With this, the drift Hamiltonian becomes
\begin{align}
    H_0 = \frac{1}{2}\sum_j \omega_j \sigma_j^z
    -J \sum_j \sigma_j^z \sigma_{j+1}^z
    -g \sum_j \sigma_j^z \sigma_{j+2}^z,
\end{align}
where $J$ is the nearest spin-spin interaction, $g=J/10$ is the next-nearest spin-spin interaction, and $\sigma$ denotes the Pauli spin operator. Here we model periodic boundary conditions and assume an isotropic spin chain ($\omega = \omega_j$ for all $j$). The control part of the Hamiltonian is
\begin{align}
    H_c(t) &= 2 \sum_{k=x,y} u^{(k)}(t) \cos(\omega t) \sum_j \sigma_j^k
    \label{eq:control_H}
\end{align}
such that $H(t) = H_0 + H_c(t)$. We may derive an effective Hamiltonian by using the rotating wave approximation (RWA)
\begin{align} \nonumber
    H_{\text{RWA}}(t) &= -J \sum_j \sigma_j^z \sigma_{j+1}^z
    -g \sum_j \sigma_j^z \sigma_{j+2}^z\\
     &+\sum_{k=x,y} u^{(k)}(t) \sum_j \sigma_j^k.
\end{align}
We express the control pulses in a modulated Fourier basis $\phi^{(k)}_n (t) $=$ s(t)\cos (\pi n t/T)$ and $\phi^{(k)}_n (t) = s(t)\sin (\pi n t/T)$ for even and odd values of $n=1,2\ldots 8$, respectively, and $k = x,y$. Here the modulation function $s(t)$ denotes a shape function that enforces the pulse to smoothly start and end at zero 
\begin{align}
    s(t) 
    =
    \begin{cases}
    &\frac{1}{2}[\cos \left(\pi (\frac{t}{\tau}-1)\right) + 1],
    \text{ for } 0 \leq t < \tau,\\
    & 1 \text{ for } \tau \leq t < T-\tau,\\
    & \frac{1}{2}[\cos \left(\pi (\frac{t-T}{\tau}+1)\right) + 1]
    \text{ for } T-\tau \leq t \leq T.\\
    \end{cases}
\end{align}
For optimization we use a sequential programming (least squares) algorithm \cite{kraft1988software} implemented in the Python library Scipy \cite{virtanen2020scipy}. This optimization algorithm allows us to handle non-linear constraints of the form
\begin{align}
    u_{\min}^{(k)} \leq u^{(k)}(t) \leq u_{\max}^{(k)} \text{  for all  }k,t,
\end{align}
with $u_{\min}^{(k)}$ and $u_{\max}^{(k)}$ denoting the minimal and maximal permissible control fields, respectively, which in this work we choose as $u_{\max}^{(k)} = -u_{\min}^{(k)} = J$ for $k=x,y$. 

We seek to make a state transfer between two degenerate ground states from an initial state $\ket{\psi_0} = \ket{0,0,\ldots,0}$ and a target state $\ket{\psi_t} = \ket{1,1,\ldots,1}$, which requires temporarily populating a set of excited states by the control fields. This task is solvable by minimizing the infidelity of the transfer  
\begin{align} 
    \mathcal{F} 
    = 1 - |\braket{\psi_t| \psi(T)}|^2, 
\end{align} 
where $\ket{\psi(T)}$ denotes the solution to the Schrödinger equation from the initial state $\ket{\psi_0} = \ket{\psi(t=0)}$. 

In the following, we compare the various integration methods for gradient-based control \rebuttal{optimization}. In order to facilitate such an analysis, we initiate a three step procedure: i) We draw $100$ seeds (initial random guesses) and optimize these using a large number of time steps, $N$, in order to estimate the “true” infidelity $\mathcal{F}_{\text{true}}$ at $TJ = 2.9$, which for the best solutions lead to an infidelity around $10^{-4}$. ii) Using the optimized pulses, we now scan over different values of $\Delta t$ in order to calculate the infidelity $\mathcal{F}_{\text{calculated}}(\Delta t)$. For each $\Delta t$ we further calculate the simulation error $|\mathcal{F}_{\text{calculated}}(\Delta t)-\mathcal{F}_{\text{true}}|$ and monitor the wall time (i.e., computational time). \rebuttal{Here the wall time is without the initialization time, which we discuss further in Sec.~\ref{sec:discussion}}. We depict in Fig.~\ref{fig:magnus_spin_chain}(a) the average simulation error as a function of the average wall time for the various methods. From the figure, we observe an excellent improvement in using the two fourth-order methods, which provides around a factor of $10$ in speed up for an average accuracy of $10^{-6}$. Interestingly, M2exact performs worse than M2approx, while M4exact performs better than M4approx. iii) At last, we also compare the various methods for control optimization. From the obtained data, we estimate for each integration scheme the $\Delta t$ that on average leads to a simulation error of $10^{-6}$. \rebuttal{This error threshold is a few orders lower than current experimental reachable infidelities and, therefore, should suffice for a fair comparison. We note that the higher-order methods generally perform better in the low-error regime, i.e., closer to machine precision.}

Using the found $\Delta t$, we then optimize the 100 seeds used in step i) again, where we now save the infidelity and wall time for each incremental update made by GRAPE. We depict the optimization trajectories (reached infidelity versus wall time) in Fig.~\ref{fig:magnus_spin_chain}(b) and Fig.~\ref{fig:magnus_spin_chain}(c) for the two approximation and the two exact methods, respectively. For an easy comparison, we further depict in Fig.~\ref{fig:magnus_spin_chain}(d) the smallest infidelity at each wall time instance for each of the different methods. \rebuttal{ As expected, the different methods converge to the same infidelity-minimum but at different speeds.} We observe around a factor of 10 speedups with the fourth-order methods compared to the second-order methods.

\subsection{Controlling a spin chain without the RWA}

\begin{figure*}
    \centering
    \includegraphics{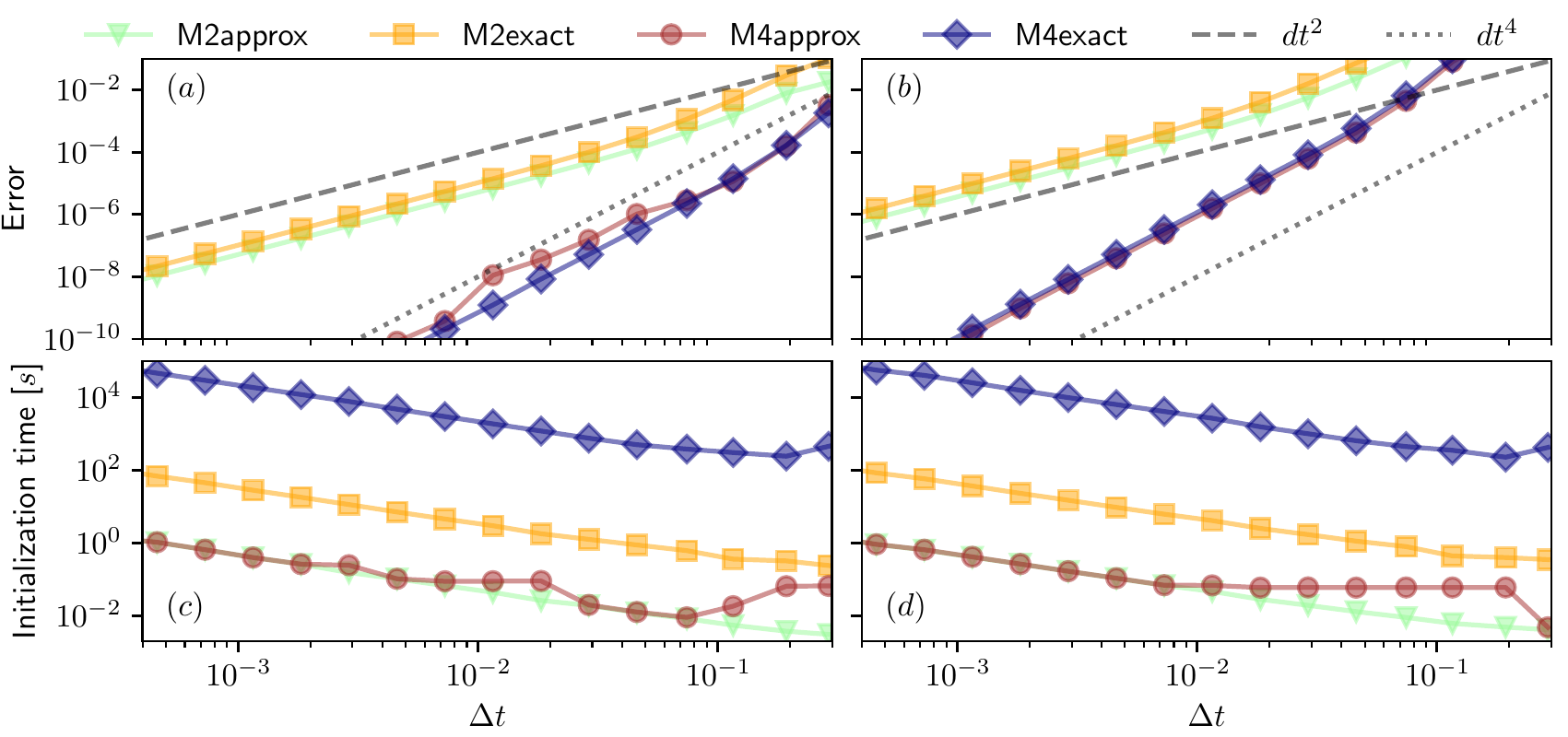}
    \caption{The error of time propagation at various values of $\Delta t$ for (a) with and (b) without the RWA. Here we observe that each method is consistent with its analytically predicted error scaling. We also depict the initialization times for (c) with and (d) without the RWA. For the two exact methods, we pre-calculate the integrals \eqref{eq:magnus_first}-\eqref{eq:third_integral} numerically and note that, where possible, calculating these analytically further reduces this initialization overhead.  }
    \label{fig:comparison_plot}
\end{figure*}

We now make an additional comparison, where we seek to simulate and optimize the spin chain without the rotating wave approximation This is easily achievable by including the cosine drives from Eq.~(\ref{eq:control_H}) into our basis functions $\phi_n^{(k)}(t) \rightarrow \phi_n^{(k)}(t) \cos (\omega t)$. Not having the rotating wave approximation requires a larger number of time steps $N$ to resolve the fast oscillations from the external fields, making simulations typically slow. In this section, we model a resonance frequency of $\omega = 20J$, which is in a regime \cite{motzoi2011optimal} where the rotating wave approximation typically leads to errors compatible with the previously optimized infidelities. In other words, this is a regime where we need the more exact model in realistic simulations. 

We now repeat the procedure from the previous section in this new scenario. In Fig.~\ref{fig:magnus_spin_chain_no_rwa}(a), we depict the simulation error versus the wall time for each of the four methods averaged over 100 optimized pulses. Overall we see the same tendencies as before: we obtain around a factor of 10 speed up with the fourth-order methods relative to the second-order ones. We still observe the counter-intuitive tendency that M2exact performs worse than M2approx. In Fig.~\ref{fig:magnus_spin_chain_no_rwa}(b) and Fig.~\ref{fig:magnus_spin_chain_no_rwa}(c) we depict the optimization trajectories for the approximate and the exact methods, respectively. Similarly to before, we also depict in Fig.~\ref{fig:magnus_spin_chain_no_rwa}(d) the smallest infidelity at each wall time instance. We again observe a factor of 10 in speed improvement with the two fourth-order methods for optimization.

\subsection{Further comparison and discussion} \label{sec:discussion}

Before concluding, we briefly elaborate on the differences between the various methods introduced and benchmarked in the preceding sections. First, we depict the aforementioned $\Delta t$-scans in Fig.~\ref{fig:comparison_plot}(a) and (b) with and without the RWA, respectively. Here we observe that each method is consistent with the second and fourth-order scaling also marked in the figures. 

We also depict the initialization time for each method in Fig.~\ref{fig:comparison_plot}(c) and (d) with and without the RWA, respectively. \rebuttal{The initialization time was not included in the wall time consumption depicted in the previous two plots, but included here for a more fair comparison.} Note that we pre-evaluate the sampling points for the two approximation methods, while for the two exact methods, we pre-calculate the integrals. In addition, we also pre-calculate the commutators for the two fourth-order methods. The most expensive method is by far M4exact. However, we believe the initialization time is heavily reducible for a large variety of control settings by pre-calculating the integrals analytically but at the price of overhead in human labor. Nonetheless, since M4exact is the consistently fastest method, there are many scenarios where this additional overhead is far smaller than the optimization time, including where many noise trajectories must be sampled, where search complexity requires many different initial seeds, or where Hilbert space complexity makes matrix exponentiation far more expensive than the integrals themselves. \rebuttal{For instance, control optimization with random sampling (as used in this work to sufficiently explore the control space) may easily take between 24 and 48 hours on a standard computer. Here, a subsequent ten-fold speed up may easily justify an additional initialization cost of one or two hours. Otherwise, M4approx avoids this high initialization cost at the price of a relatively small slow-down.}

In this work, we have only considered errors stemming from the numerical integration, which is justified by relatively low numerical errors associated with standard matrix exponentiation techniques. However, there also exists a rich class of approximations methods to the matrix exponential \cite{de2011second}, such as a second-order $\mathcal{O}(\Delta t^2)$ Suzuki-Trotter expansion \cite{ jensen2021approximate}, which may also reduce computation time with pre-calculations (e.g., eigen-decomposition of Hamiltonian terms). Thus, where Suzuki-Trotter methods are most applicable (large exponentiation cost), it may be viable to instead choose much smaller $\Delta t$ to enable their use, while having to make do with $\mathcal{O}(\Delta t^2)$ error scaling and associated slowdown.

\section{Conclusion} \label{sec:conclusion}
In this work, we have derived a new quantum optimal control framework based on improved time propagation using a discretized Magnus expansion. We obtained several high-performing quantum control propagators within this framework, which we tested in equal-accuracy comparisons to state-of-the-art approaches. In doing so, we demonstrated around an order of magnitude in speedup for both fast and slow system dynamics relative to the timescale of the control modulation. In particular, we derived an exact fourth-order Magnus propagator (M4exact) that is consistently faster than the previous methods for individual runs, though at an upfront overhead cost. At the same time, the Gauss-Legendre quadrature approximation (M4approx) offers a slight reduction in speedup while dramatically reducing this overhead. The overhead in precomputing commutators and integrals typically represents a small fraction of the entire computation time for standard optimal control problems. Still, the appropriate method can always be chosen accordingly. We foresee this new approach to have widespread applicability in optimal control-theoretic settings, and wherever many dynamical trajectories must be calculated, such as for Monte Carlo sampling.

\rebuttal{One current limitation of our framework is the difficulty in generalizing the Magnus expansion beyond the first couple of terms. One possible solution would be to combine our framework with the work of Arnal et al. \cite{arnal2018general}, which expresses the Magnus expansion in right-nested integrals and commutators. Whether or not such generalization is adaptable with our framework remains an open question and could potentially be the focus of future work. }

\acknowledgments{
We thank Francesco Preti and Michael Schilling for helpful discussions.
This work was funded by the Carlsberg Foundation, the Deutsche Forschungsgemeinschaft (DFG, German Research Foundation) under Germany's Excellence Strategy – Cluster of Excellence Matter and Light for Quantum Computing (ML4Q) EXC 2004/1 – 390534769, and through the European Union’s Horizon
2020 research and innovation programme under Grant Agreements No. 817482
(PASQuanS) and No. 820394 (ASTERIQS). The numerical results presented in this work were obtained at the Centre for Scientific Computing, Aarhus, phys.au.dk/forskning/cscaa.}

\newpage
\onecolumngrid
\appendix

\section{Different methods} \label{sec:different_methods}

\subsection{M2exact}
The first propagation scheme we consider is calculating the exact first term in the Magnus expansion. Here (and in the following methods), we start by shifting the time scale $[t_j,t_j+\Delta t]$ to $[0, \Delta t]$. In this case, we may evaluate Eq.~(\ref{eq:first_integral}) by using the parametrization $u^{k}(t) = \sum_n b_n^{(k)}\phi_n^{(k)}(t)$

\begin{align}
    c_{k,j}^{(1)} =     
    \sum_n
    b_n^{(k)}
    \int_0^{\Delta t}
    \phi_n^{(k)}(t_j + t_1)
    dt_1,
    \label{eq:first_coefficient}
\end{align}
with derivative
\begin{align}
    \frac{\partial c_{k,j}^{(1)}}{b_n^{(k)}}
    = 
    \int_0^{\Delta t}
    \phi_n^{(k)}(t_j + t_1)
    dt_1.
\end{align}
This integrals may be calculated in advance either analytically or numerically to arbitrary precision. In this work we calculated the integrals numerically. 

\subsection{M2approx (standard GRAPE)}
The second scheme we consider is the current standard in the literature: GRAPE, which is done by using midpoint interpolation of Eq.~(\ref{eq:first_coefficient}) \cite{blanes2009magnus}, which gives the coefficient
\begin{align}
    c_{k,j}^{(1)} = \Delta t \sum_n b_n^{(k)} \phi_n^{(k)}(t_j +\Delta t /2),
\end{align}
with derivative given as in \cite{motzoi2011optimal} by
\begin{align}
    \frac{\partial c_{k,j}^{(1)}}{b_n^{(k)}}
    = 
    \Delta t \phi_n^{(k)}(t_j +\Delta t /2).
\end{align}

\subsection{M4exact}
The first term in the Magnus expansion has already been calculated so we just need to evaluate Eq.~(\ref{eq:second_integral}) and (\ref{eq:third_integral}).
Then,
\begin{align}
    c_{k,j}^{(2)}
    =
    \frac{1}{2}
    \sum_n
    b_n^{(k)}
    \int_0^{\Delta t}
    \Big(
    \int_0^{t_1}
    \phi_n^{(k)}(t_j + t_2)
    dt_2
    -\phi_n^{(k)}(t_j + t_1) t_1
    \Big)
    dt_1,
\end{align}
with derivative
\begin{align}
    \frac{\partial c_{k,j}^{(2)}}{\partial b_n^{(k)}}
    =
    \frac{1}{2}
    \int_0^{\Delta t}
    \Big(
    \int_0^{t_1}
    \phi_n^{(k)}(t_j + t_2)
    dt_2
    -\phi_n^{(k)}(t_j + t_1) t_1
    \Big)
    dt_1.
\end{align}
Lastly we have
\begin{align}
    c_{k,k',j}^{(3)}
    =&
    \frac{1}{2}
    \sum_{n,m}
    \Big[
    b_n^{(k)}
    b_m^{(k')}
    \int_0^{\Delta t}
    \Big(
    \phi_n^{(k)}(t_j + t_1)
    \int_0^{t_1}
    \phi_m^{(k')}(t_j + t_2)
    dt_2
    \Big)
    dt_1\nonumber\\
    &-
    b_n^{(k')}
    b_m^{(k)}
    \int_0^{\Delta t}
    \Big(
    \phi_m^{(k')}(t_j + t_1)
    \int_0^{t_1}
    \phi_n^{(k)}(t_j + t_2)
    dt_2
    \Big)
    dt_1
    \Big],
\end{align}
with derivative
\begin{align}
    \frac{\partial c_{k,k',j}^{(3)}}{\partial b_{l}^{(h)}}
    = 
    \begin{cases}
    \frac{1}{2}
    \sum_{m}
    b_m^{(k')}
    \int_0^{\Delta t}
    \Big(
    \phi_n^{(k)}(t_j + t_1)
    \int_0^{t_1}
    \phi_m^{(k')}(t_j + t_2)
    dt_2
    \Big)
    dt_1
    &\text{ if } l=n \text{ and } h = k \\
    \frac{1}{2}
    \sum_{n}
    b_n^{(k)}
    \int_0^{\Delta t}
    \Big(
    \phi_n^{(k)}(t_j + t_1)
    \int_0^{t_1}
    \phi_m^{(k')}(t_j + t_2)
    dt_2
    \Big)
    dt_1
     &\text{ if } l=m \text{ and } h = k' \\
    -\frac{1}{2}
    \sum_{m}
    b_m^{(k)}
    \int_0^{\Delta t}
    \Big(
    \phi_m^{(k')}(t_j + t_1)
    \int_0^{t_1}
    \phi_n^{(k)}(t_j + t_2)
    dt_2
    \Big)
    dt_1
    &\text{ if } l=n \text{ and } h = k' \\
    -\frac{1}{2}
    \sum_{n}
    b_n^{(k')}
    \int_0^{\Delta t}
    \Big(
    \phi_m^{(k')}(t_j + t_1)
    \int_0^{t_1}
    \phi_n^{(k)}(t_j + t_2)
    dt_2
    \Big)
    dt_1
    &\text{ if } l=m \text{ and } h = k
    \end{cases}
\end{align}

\subsection{M4approx}
The last method we consider is an approximation scheme of fourth order based on approximating the integral with Gaussian-Legendre quadratures \cite{blanes2009magnus}. We define $H_{1,2} = H(t_j + c_{1,2} \Delta t)$ and $u_{1,2}^{(k)} = u^{(k)}( t_j + c_{1,2} \Delta t )$ where $c_{1,2} = 1/2 \mp \sqrt{3}/6$. With this the Magnus expansion reads
\begin{align}
    \Omega_j =& \frac{\Delta t }{2}
    \Big(
    H_1 + H_2
    \Big)
    -i \frac{\sqrt{3}}{12}
    \Delta t^2
    [H_2,H_1]\nonumber\\
    =&
    \Delta t H_0
    +
    \Delta t \sum_k \frac{u_1^{(k)} + u_2^{(k)} }{2}
    H_k\\
    &-i \frac{\sqrt{3}}{12} \Delta t^2
    \Big(
    \sum_{k} \big( u_1^{(k)}-u_2^{(k)}\big) [H_0, H_k]
    +
    \sum_{k<k'} 
    \big(
    u_2^{(k)} u_1^{(k')} - u_2^{(k')} u_1^{(k)}
    \big)
    [H_k, H_{k'}]
    \Big).
\end{align}
With this we have the coefficient 
\begin{align}
    c_{k,j}^{(1)}
    = \Delta t
    \frac{u_1^{(k)} + u_2^{(k)} }{2}
    = \Delta t
    \sum_n b_n^{(k)}
    \frac{
    \phi_n^{(k)}(t_1) + \phi_n^{(k)}(t_2)
    }{2},
\end{align}
where we have introduced the notation $t_{1,2} = t_j +c_{1,2} \Delta t$. The derivative is
\begin{align}
    \frac{\partial c_{k,j}^{(1)}}{b_n^{(k)}}
    =
    \Delta t
    \frac{
    \phi_n^{(k)}(t_1) + \phi_n^{(k)}(t_2)
    }{2}.
\end{align}
Then
\begin{align}
    c_{k,j}^{(2)}
    =
    \frac{\sqrt{3}}{12} \Delta t^2
    \Big(
    u_1^{(k)} - u_2^{(k)}
    \Big)
    =
    \frac{\sqrt{3}}{12} \Delta t^2
    \sum_b b_n^{(k)}
    \Big(
    \phi_n^{(k)}(t_1) - \phi_n^{(k)}(t_2)
    \Big),
\end{align}
with derivative
\begin{align}
    \frac{\partial c_{k,j}^{(2)}}{b_n^{(k)}}
    =
    \frac{\sqrt{3}}{12} \Delta t^2
    \Big(
    \phi_n^{(k)}(t_1) - \phi_n^{(k)}(t_2)
    \Big).
\end{align}
And
\begin{align}
    c_{k,k',j}^{(3)}
    &=
    \frac{\sqrt{3}}{12} \Delta t^2 
    \big(
    u_2^{(k)} u_1^{(k')}-u_2^{(k')} u_1{(k)}
    \big)\\
    &= 
    \frac{\sqrt{3}}{12} \Delta t^2
    \sum_{n,m}
    b_n^{(k)}
    b_m^{(k')}
    \phi_n^{(k)}(t_2) \phi_m^{(k')}(t_1)
    -
    b_n^{(k')}
    b_m^{(k)}
    \phi_n^{(k')}(t_2) \phi_m^{(k)}(t_1),
\end{align}
with derivative
\begin{align}
    \frac{\partial c_{k,k',j}^{(3)}}{\partial b_{l}^{(h)}}
    = 
    \begin{cases}
    \frac{\sqrt{3}}{12} \Delta t^2
    \sum_{m}
    b_m^{(k')}
    \phi_n^{(k)}(t_2) \phi_m^{(k')}(t_1)
     &\text{ if } l=n \text{ and } h = k \\
    \frac{\sqrt{3}}{12} \Delta t^2
    \sum_{n}
    b_n^{(k)}
    \phi_n^{(k)}(t_2) \phi_m^{(k')}(t_1)
     &\text{ if } l=m \text{ and } h = k' \\
    -\frac{\sqrt{3}}{12} \Delta t^2
    \sum_{m}
    b_m^{(k)}
    \phi_n^{(k')}(t_2) \phi_m^{(k)}(t_1)
    &\text{ if } l=n \text{ and } h = k' \\
    -\frac{\sqrt{3}}{12} \Delta t^2
    \sum_{n}
    b_n^{(k')}
    \phi_n^{(k')}(t_2) \phi_m^{(k)}(t_1)
    &\text{ if } l=m \text{ and } h = k
    \end{cases}
\end{align}

\twocolumngrid
\bibliography{refs}

\begin{thebibliography}{72}%
\makeatletter
\providecommand \@ifxundefined [1]{%
 \@ifx{#1\undefined}
}%
\providecommand \@ifnum [1]{%
 \ifnum #1\expandafter \@firstoftwo
 \else \expandafter \@secondoftwo
 \fi
}%
\providecommand \@ifx [1]{%
 \ifx #1\expandafter \@firstoftwo
 \else \expandafter \@secondoftwo
 \fi
}%
\providecommand \natexlab [1]{#1}%
\providecommand \enquote  [1]{``#1''}%
\providecommand \bibnamefont  [1]{#1}%
\providecommand \bibfnamefont [1]{#1}%
\providecommand \citenamefont [1]{#1}%
\providecommand \href@noop [0]{\@secondoftwo}%
\providecommand \href [0]{\begingroup \@sanitize@url \@href}%
\providecommand \@href[1]{\@@startlink{#1}\@@href}%
\providecommand \@@href[1]{\endgroup#1\@@endlink}%
\providecommand \@sanitize@url [0]{\catcode `\\12\catcode `\$12\catcode
  `\&12\catcode `\#12\catcode `\^12\catcode `\_12\catcode `\%12\relax}%
\providecommand \@@startlink[1]{}%
\providecommand \@@endlink[0]{}%
\providecommand \url  [0]{\begingroup\@sanitize@url \@url }%
\providecommand \@url [1]{\endgroup\@href {#1}{\urlprefix }}%
\providecommand \urlprefix  [0]{URL }%
\providecommand \Eprint [0]{\href }%
\providecommand \doibase [0]{http://dx.doi.org/}%
\providecommand \selectlanguage [0]{\@gobble}%
\providecommand \bibinfo  [0]{\@secondoftwo}%
\providecommand \bibfield  [0]{\@secondoftwo}%
\providecommand \translation [1]{[#1]}%
\providecommand \BibitemOpen [0]{}%
\providecommand \bibitemStop [0]{}%
\providecommand \bibitemNoStop [0]{.\EOS\space}%
\providecommand \EOS [0]{\spacefactor3000\relax}%
\providecommand \BibitemShut  [1]{\csname bibitem#1\endcsname}%
\let\auto@bib@innerbib\@empty
%</preamble>
\bibitem [{\citenamefont {Farhi}\ \emph {et~al.}(2014)\citenamefont {Farhi},
  \citenamefont {Goldstone},\ and\ \citenamefont {Gutmann}}]{farhi2014quantum}%
  \BibitemOpen
  \bibfield  {author} {\bibinfo {author} {\bibfnamefont {E.}~\bibnamefont
  {Farhi}}, \bibinfo {author} {\bibfnamefont {J.}~\bibnamefont {Goldstone}}, \
  and\ \bibinfo {author} {\bibfnamefont {S.}~\bibnamefont {Gutmann}},\
  }\href@noop {} {\bibfield  {journal} {\bibinfo  {journal} {arXiv preprint
  arXiv:1411.4028}\ } (\bibinfo {year} {2014})}\BibitemShut {NoStop}%
\bibitem [{\citenamefont {Harrow}\ \emph {et~al.}(2009)\citenamefont {Harrow},
  \citenamefont {Hassidim},\ and\ \citenamefont {Lloyd}}]{harrow2009quantum}%
  \BibitemOpen
  \bibfield  {author} {\bibinfo {author} {\bibfnamefont {A.~W.}\ \bibnamefont
  {Harrow}}, \bibinfo {author} {\bibfnamefont {A.}~\bibnamefont {Hassidim}}, \
  and\ \bibinfo {author} {\bibfnamefont {S.}~\bibnamefont {Lloyd}},\
  }\href@noop {} {\bibfield  {journal} {\bibinfo  {journal} {Phys. Rev. Lett.}\
  }\textbf {\bibinfo {volume} {103}},\ \bibinfo {pages} {150502} (\bibinfo
  {year} {2009})}\BibitemShut {NoStop}%
\bibitem [{\citenamefont {Peruzzo}\ \emph {et~al.}(2014)\citenamefont
  {Peruzzo}, \citenamefont {McClean}, \citenamefont {Shadbolt}, \citenamefont
  {Yung}, \citenamefont {Zhou}, \citenamefont {Love}, \citenamefont
  {Aspuru-Guzik},\ and\ \citenamefont {O’brien}}]{peruzzo2014variational}%
  \BibitemOpen
  \bibfield  {author} {\bibinfo {author} {\bibfnamefont {A.}~\bibnamefont
  {Peruzzo}}, \bibinfo {author} {\bibfnamefont {J.}~\bibnamefont {McClean}},
  \bibinfo {author} {\bibfnamefont {P.}~\bibnamefont {Shadbolt}}, \bibinfo
  {author} {\bibfnamefont {M.-H.}\ \bibnamefont {Yung}}, \bibinfo {author}
  {\bibfnamefont {X.-Q.}\ \bibnamefont {Zhou}}, \bibinfo {author}
  {\bibfnamefont {P.~J.}\ \bibnamefont {Love}}, \bibinfo {author}
  {\bibfnamefont {A.}~\bibnamefont {Aspuru-Guzik}}, \ and\ \bibinfo {author}
  {\bibfnamefont {J.~L.}\ \bibnamefont {O’brien}},\ }\href@noop {} {\bibfield
   {journal} {\bibinfo  {journal} {Nature communications}\ }\textbf {\bibinfo
  {volume} {5}},\ \bibinfo {pages} {1} (\bibinfo {year} {2014})}\BibitemShut
  {NoStop}%
\bibitem [{\citenamefont {McClean}\ \emph {et~al.}(2016)\citenamefont
  {McClean}, \citenamefont {Romero}, \citenamefont {Babbush},\ and\
  \citenamefont {Aspuru-Guzik}}]{mcclean2016theory}%
  \BibitemOpen
  \bibfield  {author} {\bibinfo {author} {\bibfnamefont {J.~R.}\ \bibnamefont
  {McClean}}, \bibinfo {author} {\bibfnamefont {J.}~\bibnamefont {Romero}},
  \bibinfo {author} {\bibfnamefont {R.}~\bibnamefont {Babbush}}, \ and\
  \bibinfo {author} {\bibfnamefont {A.}~\bibnamefont {Aspuru-Guzik}},\
  }\href@noop {} {\bibfield  {journal} {\bibinfo  {journal} {New J. Phys.}\
  }\textbf {\bibinfo {volume} {18}},\ \bibinfo {pages} {023023} (\bibinfo
  {year} {2016})}\BibitemShut {NoStop}%
\bibitem [{\citenamefont {Schuld}\ \emph {et~al.}(2015)\citenamefont {Schuld},
  \citenamefont {Sinayskiy},\ and\ \citenamefont
  {Petruccione}}]{schuld2015introduction}%
  \BibitemOpen
  \bibfield  {author} {\bibinfo {author} {\bibfnamefont {M.}~\bibnamefont
  {Schuld}}, \bibinfo {author} {\bibfnamefont {I.}~\bibnamefont {Sinayskiy}}, \
  and\ \bibinfo {author} {\bibfnamefont {F.}~\bibnamefont {Petruccione}},\
  }\href@noop {} {\bibfield  {journal} {\bibinfo  {journal} {Contemporary
  Physics}\ }\textbf {\bibinfo {volume} {56}},\ \bibinfo {pages} {172}
  (\bibinfo {year} {2015})}\BibitemShut {NoStop}%
\bibitem [{\citenamefont {Biamonte}\ \emph {et~al.}(2017)\citenamefont
  {Biamonte}, \citenamefont {Wittek}, \citenamefont {Pancotti}, \citenamefont
  {Rebentrost}, \citenamefont {Wiebe},\ and\ \citenamefont
  {Lloyd}}]{biamonte2017quantum}%
  \BibitemOpen
  \bibfield  {author} {\bibinfo {author} {\bibfnamefont {J.}~\bibnamefont
  {Biamonte}}, \bibinfo {author} {\bibfnamefont {P.}~\bibnamefont {Wittek}},
  \bibinfo {author} {\bibfnamefont {N.}~\bibnamefont {Pancotti}}, \bibinfo
  {author} {\bibfnamefont {P.}~\bibnamefont {Rebentrost}}, \bibinfo {author}
  {\bibfnamefont {N.}~\bibnamefont {Wiebe}}, \ and\ \bibinfo {author}
  {\bibfnamefont {S.}~\bibnamefont {Lloyd}},\ }\href@noop {} {\bibfield
  {journal} {\bibinfo  {journal} {Nature}\ }\textbf {\bibinfo {volume} {549}},\
  \bibinfo {pages} {195} (\bibinfo {year} {2017})}\BibitemShut {NoStop}%
\bibitem [{\citenamefont {Dalzell}\ \emph {et~al.}(2020)\citenamefont
  {Dalzell}, \citenamefont {Harrow}, \citenamefont {Koh},\ and\ \citenamefont
  {La~Placa}}]{dalzell2020many}%
  \BibitemOpen
  \bibfield  {author} {\bibinfo {author} {\bibfnamefont {A.~M.}\ \bibnamefont
  {Dalzell}}, \bibinfo {author} {\bibfnamefont {A.~W.}\ \bibnamefont {Harrow}},
  \bibinfo {author} {\bibfnamefont {D.~E.}\ \bibnamefont {Koh}}, \ and\
  \bibinfo {author} {\bibfnamefont {R.~L.}\ \bibnamefont {La~Placa}},\
  }\href@noop {} {\bibfield  {journal} {\bibinfo  {journal} {Quantum}\ }\textbf
  {\bibinfo {volume} {4}},\ \bibinfo {pages} {264} (\bibinfo {year}
  {2020})}\BibitemShut {NoStop}%
\bibitem [{\citenamefont {Boixo}\ \emph {et~al.}(2016)\citenamefont {Boixo},
  \citenamefont {Smelyanskiy}, \citenamefont {Shabani}, \citenamefont {Isakov},
  \citenamefont {Dykman}, \citenamefont {Denchev}, \citenamefont {Amin},
  \citenamefont {Smirnov}, \citenamefont {Mohseni},\ and\ \citenamefont
  {Neven}}]{boixo2016computational}%
  \BibitemOpen
  \bibfield  {author} {\bibinfo {author} {\bibfnamefont {S.}~\bibnamefont
  {Boixo}}, \bibinfo {author} {\bibfnamefont {V.~N.}\ \bibnamefont
  {Smelyanskiy}}, \bibinfo {author} {\bibfnamefont {A.}~\bibnamefont
  {Shabani}}, \bibinfo {author} {\bibfnamefont {S.~V.}\ \bibnamefont {Isakov}},
  \bibinfo {author} {\bibfnamefont {M.}~\bibnamefont {Dykman}}, \bibinfo
  {author} {\bibfnamefont {V.~S.}\ \bibnamefont {Denchev}}, \bibinfo {author}
  {\bibfnamefont {M.~H.}\ \bibnamefont {Amin}}, \bibinfo {author}
  {\bibfnamefont {A.~Y.}\ \bibnamefont {Smirnov}}, \bibinfo {author}
  {\bibfnamefont {M.}~\bibnamefont {Mohseni}}, \ and\ \bibinfo {author}
  {\bibfnamefont {H.}~\bibnamefont {Neven}},\ }\href@noop {} {\bibfield
  {journal} {\bibinfo  {journal} {Nature communications}\ }\textbf {\bibinfo
  {volume} {7}},\ \bibinfo {pages} {1} (\bibinfo {year} {2016})}\BibitemShut
  {NoStop}%
\bibitem [{\citenamefont {Kandala}\ \emph {et~al.}(2017)\citenamefont
  {Kandala}, \citenamefont {Mezzacapo}, \citenamefont {Temme}, \citenamefont
  {Takita}, \citenamefont {Brink}, \citenamefont {Chow},\ and\ \citenamefont
  {Gambetta}}]{kandala2017hardware}%
  \BibitemOpen
  \bibfield  {author} {\bibinfo {author} {\bibfnamefont {A.}~\bibnamefont
  {Kandala}}, \bibinfo {author} {\bibfnamefont {A.}~\bibnamefont {Mezzacapo}},
  \bibinfo {author} {\bibfnamefont {K.}~\bibnamefont {Temme}}, \bibinfo
  {author} {\bibfnamefont {M.}~\bibnamefont {Takita}}, \bibinfo {author}
  {\bibfnamefont {M.}~\bibnamefont {Brink}}, \bibinfo {author} {\bibfnamefont
  {J.~M.}\ \bibnamefont {Chow}}, \ and\ \bibinfo {author} {\bibfnamefont
  {J.~M.}\ \bibnamefont {Gambetta}},\ }\href@noop {} {\bibfield  {journal}
  {\bibinfo  {journal} {Nature}\ }\textbf {\bibinfo {volume} {549}},\ \bibinfo
  {pages} {242} (\bibinfo {year} {2017})}\BibitemShut {NoStop}%
\bibitem [{\citenamefont {Kokail}\ \emph {et~al.}(2019)\citenamefont {Kokail},
  \citenamefont {Maier}, \citenamefont {van Bijnen}, \citenamefont {Brydges},
  \citenamefont {Joshi}, \citenamefont {Jurcevic}, \citenamefont {Muschik},
  \citenamefont {Silvi}, \citenamefont {Blatt}, \citenamefont {Roos} \emph
  {et~al.}}]{kokail2019self}%
  \BibitemOpen
  \bibfield  {author} {\bibinfo {author} {\bibfnamefont {C.}~\bibnamefont
  {Kokail}}, \bibinfo {author} {\bibfnamefont {C.}~\bibnamefont {Maier}},
  \bibinfo {author} {\bibfnamefont {R.}~\bibnamefont {van Bijnen}}, \bibinfo
  {author} {\bibfnamefont {T.}~\bibnamefont {Brydges}}, \bibinfo {author}
  {\bibfnamefont {M.~K.}\ \bibnamefont {Joshi}}, \bibinfo {author}
  {\bibfnamefont {P.}~\bibnamefont {Jurcevic}}, \bibinfo {author}
  {\bibfnamefont {C.~A.}\ \bibnamefont {Muschik}}, \bibinfo {author}
  {\bibfnamefont {P.}~\bibnamefont {Silvi}}, \bibinfo {author} {\bibfnamefont
  {R.}~\bibnamefont {Blatt}}, \bibinfo {author} {\bibfnamefont {C.~F.}\
  \bibnamefont {Roos}},  \emph {et~al.},\ }\href@noop {} {\bibfield  {journal}
  {\bibinfo  {journal} {Nature}\ }\textbf {\bibinfo {volume} {569}},\ \bibinfo
  {pages} {355} (\bibinfo {year} {2019})}\BibitemShut {NoStop}%
\bibitem [{\citenamefont {Omran}\ \emph {et~al.}(2019)\citenamefont {Omran},
  \citenamefont {Levine}, \citenamefont {Keesling}, \citenamefont {Semeghini},
  \citenamefont {Wang}, \citenamefont {Ebadi}, \citenamefont {Bernien},
  \citenamefont {Zibrov}, \citenamefont {Pichler}, \citenamefont {Choi} \emph
  {et~al.}}]{omran2019generation}%
  \BibitemOpen
  \bibfield  {author} {\bibinfo {author} {\bibfnamefont {A.}~\bibnamefont
  {Omran}}, \bibinfo {author} {\bibfnamefont {H.}~\bibnamefont {Levine}},
  \bibinfo {author} {\bibfnamefont {A.}~\bibnamefont {Keesling}}, \bibinfo
  {author} {\bibfnamefont {G.}~\bibnamefont {Semeghini}}, \bibinfo {author}
  {\bibfnamefont {T.~T.}\ \bibnamefont {Wang}}, \bibinfo {author}
  {\bibfnamefont {S.}~\bibnamefont {Ebadi}}, \bibinfo {author} {\bibfnamefont
  {H.}~\bibnamefont {Bernien}}, \bibinfo {author} {\bibfnamefont {A.~S.}\
  \bibnamefont {Zibrov}}, \bibinfo {author} {\bibfnamefont {H.}~\bibnamefont
  {Pichler}}, \bibinfo {author} {\bibfnamefont {S.}~\bibnamefont {Choi}},
  \emph {et~al.},\ }\href@noop {} {\bibfield  {journal} {\bibinfo  {journal}
  {Science}\ }\textbf {\bibinfo {volume} {365}},\ \bibinfo {pages} {570}
  (\bibinfo {year} {2019})}\BibitemShut {NoStop}%
\bibitem [{\citenamefont {Harrigan}\ \emph {et~al.}(2021)\citenamefont
  {Harrigan}, \citenamefont {Sung}, \citenamefont {Neeley}, \citenamefont
  {Satzinger}, \citenamefont {Arute}, \citenamefont {Arya}, \citenamefont
  {Atalaya}, \citenamefont {Bardin}, \citenamefont {Barends}, \citenamefont
  {Boixo} \emph {et~al.}}]{harrigan2021quantum}%
  \BibitemOpen
  \bibfield  {author} {\bibinfo {author} {\bibfnamefont {M.~P.}\ \bibnamefont
  {Harrigan}}, \bibinfo {author} {\bibfnamefont {K.~J.}\ \bibnamefont {Sung}},
  \bibinfo {author} {\bibfnamefont {M.}~\bibnamefont {Neeley}}, \bibinfo
  {author} {\bibfnamefont {K.~J.}\ \bibnamefont {Satzinger}}, \bibinfo {author}
  {\bibfnamefont {F.}~\bibnamefont {Arute}}, \bibinfo {author} {\bibfnamefont
  {K.}~\bibnamefont {Arya}}, \bibinfo {author} {\bibfnamefont {J.}~\bibnamefont
  {Atalaya}}, \bibinfo {author} {\bibfnamefont {J.~C.}\ \bibnamefont {Bardin}},
  \bibinfo {author} {\bibfnamefont {R.}~\bibnamefont {Barends}}, \bibinfo
  {author} {\bibfnamefont {S.}~\bibnamefont {Boixo}},  \emph {et~al.},\
  }\href@noop {} {\bibfield  {journal} {\bibinfo  {journal} {Nature Physics}\
  }\textbf {\bibinfo {volume} {17}},\ \bibinfo {pages} {332} (\bibinfo {year}
  {2021})}\BibitemShut {NoStop}%
\bibitem [{\citenamefont {d'Alessandro}(2007)}]{d2007introduction}%
  \BibitemOpen
  \bibfield  {author} {\bibinfo {author} {\bibfnamefont {D.}~\bibnamefont
  {d'Alessandro}},\ }\href@noop {} {\emph {\bibinfo {title} {Introduction to
  quantum control and dynamics}}}\ (\bibinfo  {publisher} {Chapman and
  Hall/CRC},\ \bibinfo {year} {2007})\BibitemShut {NoStop}%
\bibitem [{\citenamefont {Werschnik}\ and\ \citenamefont
  {Gross}(2007)}]{werschnik2007quantum}%
  \BibitemOpen
  \bibfield  {author} {\bibinfo {author} {\bibfnamefont {J.}~\bibnamefont
  {Werschnik}}\ and\ \bibinfo {author} {\bibfnamefont {E.}~\bibnamefont
  {Gross}},\ }\href@noop {} {\bibfield  {journal} {\bibinfo  {journal} {Journal
  of Physics B: Atomic, Molecular and Optical Physics}\ }\textbf {\bibinfo
  {volume} {40}},\ \bibinfo {pages} {R175} (\bibinfo {year}
  {2007})}\BibitemShut {NoStop}%
\bibitem [{\citenamefont {Glaser}\ \emph {et~al.}(2015)\citenamefont {Glaser},
  \citenamefont {Boscain}, \citenamefont {Calarco}, \citenamefont {Koch},
  \citenamefont {K{\"o}ckenberger}, \citenamefont {Kosloff}, \citenamefont
  {Kuprov}, \citenamefont {Luy}, \citenamefont {Schirmer}, \citenamefont
  {Schulte-Herbr{\"u}ggen} \emph {et~al.}}]{glaser2015training}%
  \BibitemOpen
  \bibfield  {author} {\bibinfo {author} {\bibfnamefont {S.~J.}\ \bibnamefont
  {Glaser}}, \bibinfo {author} {\bibfnamefont {U.}~\bibnamefont {Boscain}},
  \bibinfo {author} {\bibfnamefont {T.}~\bibnamefont {Calarco}}, \bibinfo
  {author} {\bibfnamefont {C.~P.}\ \bibnamefont {Koch}}, \bibinfo {author}
  {\bibfnamefont {W.}~\bibnamefont {K{\"o}ckenberger}}, \bibinfo {author}
  {\bibfnamefont {R.}~\bibnamefont {Kosloff}}, \bibinfo {author} {\bibfnamefont
  {I.}~\bibnamefont {Kuprov}}, \bibinfo {author} {\bibfnamefont
  {B.}~\bibnamefont {Luy}}, \bibinfo {author} {\bibfnamefont {S.}~\bibnamefont
  {Schirmer}}, \bibinfo {author} {\bibfnamefont {T.}~\bibnamefont
  {Schulte-Herbr{\"u}ggen}},  \emph {et~al.},\ }\href@noop {} {\bibfield
  {journal} {\bibinfo  {journal} {The European Physical Journal D}\ }\textbf
  {\bibinfo {volume} {69}},\ \bibinfo {pages} {1} (\bibinfo {year}
  {2015})}\BibitemShut {NoStop}%
\bibitem [{\citenamefont {Goerz}(2015)}]{Georzthesis}%
  \BibitemOpen
  \bibfield  {author} {\bibinfo {author} {\bibfnamefont {M.~H.}\ \bibnamefont
  {Goerz}},\ }\emph {\bibinfo {title} {Optimizing Robust Quantum Gates in Open
  Quantum Systems}},\ \href@noop {} {Ph.D. thesis},\ \bibinfo  {school}
  {Universität Kassel} (\bibinfo {year} {2015})\BibitemShut {NoStop}%
\bibitem [{\citenamefont {Dong}\ \emph {et~al.}(2020)\citenamefont {Dong},
  \citenamefont {Shu}, \citenamefont {Chen}, \citenamefont {Xing},
  \citenamefont {Ma}, \citenamefont {Guo},\ and\ \citenamefont
  {Rabitz}}]{dong2020learning}%
  \BibitemOpen
  \bibfield  {author} {\bibinfo {author} {\bibfnamefont {D.}~\bibnamefont
  {Dong}}, \bibinfo {author} {\bibfnamefont {C.-C.}\ \bibnamefont {Shu}},
  \bibinfo {author} {\bibfnamefont {J.}~\bibnamefont {Chen}}, \bibinfo {author}
  {\bibfnamefont {X.}~\bibnamefont {Xing}}, \bibinfo {author} {\bibfnamefont
  {H.}~\bibnamefont {Ma}}, \bibinfo {author} {\bibfnamefont {Y.}~\bibnamefont
  {Guo}}, \ and\ \bibinfo {author} {\bibfnamefont {H.}~\bibnamefont {Rabitz}},\
  }\href@noop {} {\bibfield  {journal} {\bibinfo  {journal} {IEEE Transactions
  on Control Systems Technology}\ } (\bibinfo {year} {2020})}\BibitemShut
  {NoStop}%
\bibitem [{\citenamefont {S{\o}rensen}\ \emph {et~al.}(2020)\citenamefont
  {S{\o}rensen}, \citenamefont {Nyemann}, \citenamefont {Motzoi}, \citenamefont
  {Sherson},\ and\ \citenamefont {Vosegaard}}]{sorensen2020optimization}%
  \BibitemOpen
  \bibfield  {author} {\bibinfo {author} {\bibfnamefont {J.~J.}\ \bibnamefont
  {S{\o}rensen}}, \bibinfo {author} {\bibfnamefont {J.~S.}\ \bibnamefont
  {Nyemann}}, \bibinfo {author} {\bibfnamefont {F.}~\bibnamefont {Motzoi}},
  \bibinfo {author} {\bibfnamefont {J.}~\bibnamefont {Sherson}}, \ and\
  \bibinfo {author} {\bibfnamefont {T.}~\bibnamefont {Vosegaard}},\ }\href@noop
  {} {\bibfield  {journal} {\bibinfo  {journal} {The Journal of Chemical
  Physics}\ }\textbf {\bibinfo {volume} {152}},\ \bibinfo {pages} {054104}
  (\bibinfo {year} {2020})}\BibitemShut {NoStop}%
\bibitem [{\citenamefont {M{\"u}ller}\ \emph {et~al.}(2021)\citenamefont
  {M{\"u}ller}, \citenamefont {Said}, \citenamefont {Jelezko}, \citenamefont
  {Calarco},\ and\ \citenamefont {Montangero}}]{muller2021one}%
  \BibitemOpen
  \bibfield  {author} {\bibinfo {author} {\bibfnamefont {M.~M.}\ \bibnamefont
  {M{\"u}ller}}, \bibinfo {author} {\bibfnamefont {R.~S.}\ \bibnamefont
  {Said}}, \bibinfo {author} {\bibfnamefont {F.}~\bibnamefont {Jelezko}},
  \bibinfo {author} {\bibfnamefont {T.}~\bibnamefont {Calarco}}, \ and\
  \bibinfo {author} {\bibfnamefont {S.}~\bibnamefont {Montangero}},\
  }\href@noop {} {\bibfield  {journal} {\bibinfo  {journal} {arXiv preprint
  arXiv:2104.07687}\ } (\bibinfo {year} {2021})}\BibitemShut {NoStop}%
\bibitem [{\citenamefont {Goodwin}\ and\ \citenamefont
  {Kuprov}(2016)}]{goodwin2016modified}%
  \BibitemOpen
  \bibfield  {author} {\bibinfo {author} {\bibfnamefont {D.}~\bibnamefont
  {Goodwin}}\ and\ \bibinfo {author} {\bibfnamefont {I.}~\bibnamefont
  {Kuprov}},\ }\href@noop {} {\bibfield  {journal} {\bibinfo  {journal} {The
  Journal of chemical physics}\ }\textbf {\bibinfo {volume} {144}},\ \bibinfo
  {pages} {204107} (\bibinfo {year} {2016})}\BibitemShut {NoStop}%
\bibitem [{\citenamefont {Dalgaard}\ \emph
  {et~al.}(2020{\natexlab{a}})\citenamefont {Dalgaard}, \citenamefont {Motzoi},
  \citenamefont {Jensen},\ and\ \citenamefont {Sherson}}]{dalgaard2020hessian}%
  \BibitemOpen
  \bibfield  {author} {\bibinfo {author} {\bibfnamefont {M.}~\bibnamefont
  {Dalgaard}}, \bibinfo {author} {\bibfnamefont {F.}~\bibnamefont {Motzoi}},
  \bibinfo {author} {\bibfnamefont {J.~H.~M.}\ \bibnamefont {Jensen}}, \ and\
  \bibinfo {author} {\bibfnamefont {J.}~\bibnamefont {Sherson}},\ }\href@noop
  {} {\bibfield  {journal} {\bibinfo  {journal} {Phys. Rev. A}\ }\textbf
  {\bibinfo {volume} {102}},\ \bibinfo {pages} {042612} (\bibinfo {year}
  {2020}{\natexlab{a}})}\BibitemShut {NoStop}%
\bibitem [{\citenamefont {Jensen}\ \emph {et~al.}(2021)\citenamefont {Jensen},
  \citenamefont {M{\o}ller}, \citenamefont {S{\o}rensen},\ and\ \citenamefont
  {Sherson}}]{jensen2021approximate}%
  \BibitemOpen
  \bibfield  {author} {\bibinfo {author} {\bibfnamefont {J.~H.~M.}\
  \bibnamefont {Jensen}}, \bibinfo {author} {\bibfnamefont {F.~S.}\
  \bibnamefont {M{\o}ller}}, \bibinfo {author} {\bibfnamefont {J.~J.}\
  \bibnamefont {S{\o}rensen}}, \ and\ \bibinfo {author} {\bibfnamefont {J.~F.}\
  \bibnamefont {Sherson}},\ }\href@noop {} {\bibfield  {journal} {\bibinfo
  {journal} {Phys. Rev. A}\ }\textbf {\bibinfo {volume} {103}},\ \bibinfo
  {pages} {062612} (\bibinfo {year} {2021})}\BibitemShut {NoStop}%
\bibitem [{\citenamefont {Jensen}\ \emph {et~al.}(2020)\citenamefont {Jensen},
  \citenamefont {M{\o}ller}, \citenamefont {S{\o}rensen},\ and\ \citenamefont
  {Sherson}}]{jensen2020achieving}%
  \BibitemOpen
  \bibfield  {author} {\bibinfo {author} {\bibfnamefont {J.~H.~M.}\
  \bibnamefont {Jensen}}, \bibinfo {author} {\bibfnamefont {F.~S.}\
  \bibnamefont {M{\o}ller}}, \bibinfo {author} {\bibfnamefont {J.~J.}\
  \bibnamefont {S{\o}rensen}}, \ and\ \bibinfo {author} {\bibfnamefont {J.~F.}\
  \bibnamefont {Sherson}},\ }\href@noop {} {\bibfield  {journal} {\bibinfo
  {journal} {arXiv preprint arXiv:2008.06076}\ } (\bibinfo {year}
  {2020})}\BibitemShut {NoStop}%
\bibitem [{\citenamefont {M{\"u}ller}\ \emph {et~al.}(2018)\citenamefont
  {M{\"u}ller}, \citenamefont {Gherardini},\ and\ \citenamefont
  {Caruso}}]{muller2018noise}%
  \BibitemOpen
  \bibfield  {author} {\bibinfo {author} {\bibfnamefont {M.~M.}\ \bibnamefont
  {M{\"u}ller}}, \bibinfo {author} {\bibfnamefont {S.}~\bibnamefont
  {Gherardini}}, \ and\ \bibinfo {author} {\bibfnamefont {F.}~\bibnamefont
  {Caruso}},\ }\href@noop {} {\bibfield  {journal} {\bibinfo  {journal}
  {Scientific reports}\ }\textbf {\bibinfo {volume} {8}},\ \bibinfo {pages} {1}
  (\bibinfo {year} {2018})}\BibitemShut {NoStop}%
\bibitem [{\citenamefont {Liu}\ \emph {et~al.}(2019)\citenamefont {Liu},
  \citenamefont {Song}, \citenamefont {Xue}, \citenamefont {Wang},\ and\
  \citenamefont {Yung}}]{liu2019plug}%
  \BibitemOpen
  \bibfield  {author} {\bibinfo {author} {\bibfnamefont {B.-J.}\ \bibnamefont
  {Liu}}, \bibinfo {author} {\bibfnamefont {X.-K.}\ \bibnamefont {Song}},
  \bibinfo {author} {\bibfnamefont {Z.-Y.}\ \bibnamefont {Xue}}, \bibinfo
  {author} {\bibfnamefont {X.}~\bibnamefont {Wang}}, \ and\ \bibinfo {author}
  {\bibfnamefont {M.-H.}\ \bibnamefont {Yung}},\ }\href@noop {} {\bibfield
  {journal} {\bibinfo  {journal} {Phys. Rev. Lett.}\ }\textbf {\bibinfo
  {volume} {123}},\ \bibinfo {pages} {100501} (\bibinfo {year}
  {2019})}\BibitemShut {NoStop}%
\bibitem [{\citenamefont {Khani}\ \emph {et~al.}(2012)\citenamefont {Khani},
  \citenamefont {Merkel}, \citenamefont {Motzoi}, \citenamefont {Gambetta},\
  and\ \citenamefont {Wilhelm}}]{khani2012high}%
  \BibitemOpen
  \bibfield  {author} {\bibinfo {author} {\bibfnamefont {B.}~\bibnamefont
  {Khani}}, \bibinfo {author} {\bibfnamefont {S.~T.}\ \bibnamefont {Merkel}},
  \bibinfo {author} {\bibfnamefont {F.}~\bibnamefont {Motzoi}}, \bibinfo
  {author} {\bibfnamefont {J.~M.}\ \bibnamefont {Gambetta}}, \ and\ \bibinfo
  {author} {\bibfnamefont {F.~K.}\ \bibnamefont {Wilhelm}},\ }\href@noop {}
  {\bibfield  {journal} {\bibinfo  {journal} {Phys. Rev. A}\ }\textbf {\bibinfo
  {volume} {85}},\ \bibinfo {pages} {022306} (\bibinfo {year}
  {2012})}\BibitemShut {NoStop}%
\bibitem [{\citenamefont {Gupta}\ \emph {et~al.}(2020)\citenamefont {Gupta},
  \citenamefont {Edmunds}, \citenamefont {Milne}, \citenamefont {Hempel},\ and\
  \citenamefont {Biercuk}}]{gupta2020adaptive}%
  \BibitemOpen
  \bibfield  {author} {\bibinfo {author} {\bibfnamefont {R.~S.}\ \bibnamefont
  {Gupta}}, \bibinfo {author} {\bibfnamefont {C.~L.}\ \bibnamefont {Edmunds}},
  \bibinfo {author} {\bibfnamefont {A.~R.}\ \bibnamefont {Milne}}, \bibinfo
  {author} {\bibfnamefont {C.}~\bibnamefont {Hempel}}, \ and\ \bibinfo {author}
  {\bibfnamefont {M.~J.}\ \bibnamefont {Biercuk}},\ }\href@noop {} {\bibfield
  {journal} {\bibinfo  {journal} {npj Quantum Information}\ }\textbf {\bibinfo
  {volume} {6}},\ \bibinfo {pages} {1} (\bibinfo {year} {2020})}\BibitemShut
  {NoStop}%
\bibitem [{\citenamefont {Dalgaard}\ \emph
  {et~al.}(2021{\natexlab{a}})\citenamefont {Dalgaard}, \citenamefont
  {Weidner},\ and\ \citenamefont {Motzoi}}]{dalgaard2021dynamical}%
  \BibitemOpen
  \bibfield  {author} {\bibinfo {author} {\bibfnamefont {M.}~\bibnamefont
  {Dalgaard}}, \bibinfo {author} {\bibfnamefont {C.~A.}\ \bibnamefont
  {Weidner}}, \ and\ \bibinfo {author} {\bibfnamefont {F.}~\bibnamefont
  {Motzoi}},\ }\href@noop {} {\bibfield  {journal} {\bibinfo  {journal} {arXiv
  preprint arXiv:2107.11388}\ } (\bibinfo {year}
  {2021}{\natexlab{a}})}\BibitemShut {NoStop}%
\bibitem [{\citenamefont {Palittapongarnpim}\ \emph {et~al.}(2017)\citenamefont
  {Palittapongarnpim}, \citenamefont {Wittek}, \citenamefont {Zahedinejad},
  \citenamefont {Vedaie},\ and\ \citenamefont
  {Sanders}}]{palittapongarnpim2017learning}%
  \BibitemOpen
  \bibfield  {author} {\bibinfo {author} {\bibfnamefont {P.}~\bibnamefont
  {Palittapongarnpim}}, \bibinfo {author} {\bibfnamefont {P.}~\bibnamefont
  {Wittek}}, \bibinfo {author} {\bibfnamefont {E.}~\bibnamefont {Zahedinejad}},
  \bibinfo {author} {\bibfnamefont {S.}~\bibnamefont {Vedaie}}, \ and\ \bibinfo
  {author} {\bibfnamefont {B.~C.}\ \bibnamefont {Sanders}},\ }\href@noop {}
  {\bibfield  {journal} {\bibinfo  {journal} {Neurocomputing}\ }\textbf
  {\bibinfo {volume} {268}},\ \bibinfo {pages} {116} (\bibinfo {year}
  {2017})}\BibitemShut {NoStop}%
\bibitem [{\citenamefont {Bukov}\ \emph {et~al.}(2018)\citenamefont {Bukov},
  \citenamefont {Day}, \citenamefont {Sels}, \citenamefont {Weinberg},
  \citenamefont {Polkovnikov},\ and\ \citenamefont
  {Mehta}}]{bukov2018reinforcement}%
  \BibitemOpen
  \bibfield  {author} {\bibinfo {author} {\bibfnamefont {M.}~\bibnamefont
  {Bukov}}, \bibinfo {author} {\bibfnamefont {A.~G.}\ \bibnamefont {Day}},
  \bibinfo {author} {\bibfnamefont {D.}~\bibnamefont {Sels}}, \bibinfo {author}
  {\bibfnamefont {P.}~\bibnamefont {Weinberg}}, \bibinfo {author}
  {\bibfnamefont {A.}~\bibnamefont {Polkovnikov}}, \ and\ \bibinfo {author}
  {\bibfnamefont {P.}~\bibnamefont {Mehta}},\ }\href@noop {} {\bibfield
  {journal} {\bibinfo  {journal} {Phys. Rev. X}\ }\textbf {\bibinfo {volume}
  {8}},\ \bibinfo {pages} {031086} (\bibinfo {year} {2018})}\BibitemShut
  {NoStop}%
\bibitem [{\citenamefont {Xu}\ \emph {et~al.}(2019)\citenamefont {Xu},
  \citenamefont {Li}, \citenamefont {Liu}, \citenamefont {Wang}, \citenamefont
  {Yuan},\ and\ \citenamefont {Wang}}]{xu2019generalizable}%
  \BibitemOpen
  \bibfield  {author} {\bibinfo {author} {\bibfnamefont {H.}~\bibnamefont
  {Xu}}, \bibinfo {author} {\bibfnamefont {J.}~\bibnamefont {Li}}, \bibinfo
  {author} {\bibfnamefont {L.}~\bibnamefont {Liu}}, \bibinfo {author}
  {\bibfnamefont {Y.}~\bibnamefont {Wang}}, \bibinfo {author} {\bibfnamefont
  {H.}~\bibnamefont {Yuan}}, \ and\ \bibinfo {author} {\bibfnamefont
  {X.}~\bibnamefont {Wang}},\ }\href@noop {} {\bibfield  {journal} {\bibinfo
  {journal} {npj Quantum Information}\ }\textbf {\bibinfo {volume} {5}},\
  \bibinfo {pages} {1} (\bibinfo {year} {2019})}\BibitemShut {NoStop}%
\bibitem [{\citenamefont {An}\ and\ \citenamefont {Zhou}(2019)}]{an2019deep}%
  \BibitemOpen
  \bibfield  {author} {\bibinfo {author} {\bibfnamefont {Z.}~\bibnamefont
  {An}}\ and\ \bibinfo {author} {\bibfnamefont {D.}~\bibnamefont {Zhou}},\
  }\href@noop {} {\bibfield  {journal} {\bibinfo  {journal} {EPL (Europhysics
  Letters)}\ }\textbf {\bibinfo {volume} {126}},\ \bibinfo {pages} {60002}
  (\bibinfo {year} {2019})}\BibitemShut {NoStop}%
\bibitem [{\citenamefont {Dalgaard}\ \emph
  {et~al.}(2020{\natexlab{b}})\citenamefont {Dalgaard}, \citenamefont {Motzoi},
  \citenamefont {S{\o}rensen},\ and\ \citenamefont
  {Sherson}}]{dalgaard2020global}%
  \BibitemOpen
  \bibfield  {author} {\bibinfo {author} {\bibfnamefont {M.}~\bibnamefont
  {Dalgaard}}, \bibinfo {author} {\bibfnamefont {F.}~\bibnamefont {Motzoi}},
  \bibinfo {author} {\bibfnamefont {J.~J.}\ \bibnamefont {S{\o}rensen}}, \ and\
  \bibinfo {author} {\bibfnamefont {J.}~\bibnamefont {Sherson}},\ }\href@noop
  {} {\bibfield  {journal} {\bibinfo  {journal} {npj Quantum Information}\
  }\textbf {\bibinfo {volume} {6}},\ \bibinfo {pages} {1} (\bibinfo {year}
  {2020}{\natexlab{b}})}\BibitemShut {NoStop}%
\bibitem [{\citenamefont {Yao}\ \emph {et~al.}(2020)\citenamefont {Yao},
  \citenamefont {Bukov},\ and\ \citenamefont {Lin}}]{yao2020policy}%
  \BibitemOpen
  \bibfield  {author} {\bibinfo {author} {\bibfnamefont {J.}~\bibnamefont
  {Yao}}, \bibinfo {author} {\bibfnamefont {M.}~\bibnamefont {Bukov}}, \ and\
  \bibinfo {author} {\bibfnamefont {L.}~\bibnamefont {Lin}},\ }in\ \href@noop
  {} {\emph {\bibinfo {booktitle} {Mathematical and Scientific Machine
  Learning}}}\ (\bibinfo {organization} {PMLR},\ \bibinfo {year} {2020})\ pp.\
  \bibinfo {pages} {605--634}\BibitemShut {NoStop}%
\bibitem [{\citenamefont {Baum}\ \emph {et~al.}(2021)\citenamefont {Baum},
  \citenamefont {Amico}, \citenamefont {Howell}, \citenamefont {Hush},
  \citenamefont {Liuzzi}, \citenamefont {Mundada}, \citenamefont {Merkh},
  \citenamefont {Carvalho},\ and\ \citenamefont
  {Biercuk}}]{baum2021experimental}%
  \BibitemOpen
  \bibfield  {author} {\bibinfo {author} {\bibfnamefont {Y.}~\bibnamefont
  {Baum}}, \bibinfo {author} {\bibfnamefont {M.}~\bibnamefont {Amico}},
  \bibinfo {author} {\bibfnamefont {S.}~\bibnamefont {Howell}}, \bibinfo
  {author} {\bibfnamefont {M.}~\bibnamefont {Hush}}, \bibinfo {author}
  {\bibfnamefont {M.}~\bibnamefont {Liuzzi}}, \bibinfo {author} {\bibfnamefont
  {P.}~\bibnamefont {Mundada}}, \bibinfo {author} {\bibfnamefont
  {T.}~\bibnamefont {Merkh}}, \bibinfo {author} {\bibfnamefont {A.~R.}\
  \bibnamefont {Carvalho}}, \ and\ \bibinfo {author} {\bibfnamefont {M.~J.}\
  \bibnamefont {Biercuk}},\ }\href@noop {} {\bibfield  {journal} {\bibinfo
  {journal} {arXiv preprint arXiv:2105.01079}\ } (\bibinfo {year}
  {2021})}\BibitemShut {NoStop}%
\bibitem [{\citenamefont {Mitarai}\ \emph {et~al.}(2018)\citenamefont
  {Mitarai}, \citenamefont {Negoro}, \citenamefont {Kitagawa},\ and\
  \citenamefont {Fujii}}]{mitarai2018quantum}%
  \BibitemOpen
  \bibfield  {author} {\bibinfo {author} {\bibfnamefont {K.}~\bibnamefont
  {Mitarai}}, \bibinfo {author} {\bibfnamefont {M.}~\bibnamefont {Negoro}},
  \bibinfo {author} {\bibfnamefont {M.}~\bibnamefont {Kitagawa}}, \ and\
  \bibinfo {author} {\bibfnamefont {K.}~\bibnamefont {Fujii}},\ }\href@noop {}
  {\bibfield  {journal} {\bibinfo  {journal} {Phys. Rev. A}\ }\textbf {\bibinfo
  {volume} {98}},\ \bibinfo {pages} {032309} (\bibinfo {year}
  {2018})}\BibitemShut {NoStop}%
\bibitem [{\citenamefont {Motzoi}\ \emph {et~al.}(2017)\citenamefont {Motzoi},
  \citenamefont {Kaicher},\ and\ \citenamefont {Wilhelm}}]{motzoi2017linear}%
  \BibitemOpen
  \bibfield  {author} {\bibinfo {author} {\bibfnamefont {F.}~\bibnamefont
  {Motzoi}}, \bibinfo {author} {\bibfnamefont {M.~P.}\ \bibnamefont {Kaicher}},
  \ and\ \bibinfo {author} {\bibfnamefont {F.~K.}\ \bibnamefont {Wilhelm}},\
  }\href@noop {} {\bibfield  {journal} {\bibinfo  {journal} {Phys. Rev. Lett.}\
  }\textbf {\bibinfo {volume} {119}},\ \bibinfo {pages} {160503} (\bibinfo
  {year} {2017})}\BibitemShut {NoStop}%
\bibitem [{\citenamefont {Arrazola}\ \emph {et~al.}(2019)\citenamefont
  {Arrazola}, \citenamefont {Bromley}, \citenamefont {Izaac}, \citenamefont
  {Myers}, \citenamefont {Br{\'a}dler},\ and\ \citenamefont
  {Killoran}}]{arrazola2019machine}%
  \BibitemOpen
  \bibfield  {author} {\bibinfo {author} {\bibfnamefont {J.~M.}\ \bibnamefont
  {Arrazola}}, \bibinfo {author} {\bibfnamefont {T.~R.}\ \bibnamefont
  {Bromley}}, \bibinfo {author} {\bibfnamefont {J.}~\bibnamefont {Izaac}},
  \bibinfo {author} {\bibfnamefont {C.~R.}\ \bibnamefont {Myers}}, \bibinfo
  {author} {\bibfnamefont {K.}~\bibnamefont {Br{\'a}dler}}, \ and\ \bibinfo
  {author} {\bibfnamefont {N.}~\bibnamefont {Killoran}},\ }\href@noop {}
  {\bibfield  {journal} {\bibinfo  {journal} {Quantum Science and Technology}\
  }\textbf {\bibinfo {volume} {4}},\ \bibinfo {pages} {024004} (\bibinfo {year}
  {2019})}\BibitemShut {NoStop}%
\bibitem [{\citenamefont {Chen}\ \emph {et~al.}(2020)\citenamefont {Chen},
  \citenamefont {Li}, \citenamefont {Motzoi}, \citenamefont {Martin},
  \citenamefont {Whaley},\ and\ \citenamefont {Sarovar}}]{chen2020quantum}%
  \BibitemOpen
  \bibfield  {author} {\bibinfo {author} {\bibfnamefont {H.}~\bibnamefont
  {Chen}}, \bibinfo {author} {\bibfnamefont {H.}~\bibnamefont {Li}}, \bibinfo
  {author} {\bibfnamefont {F.}~\bibnamefont {Motzoi}}, \bibinfo {author}
  {\bibfnamefont {L.}~\bibnamefont {Martin}}, \bibinfo {author} {\bibfnamefont
  {K.~B.}\ \bibnamefont {Whaley}}, \ and\ \bibinfo {author} {\bibfnamefont
  {M.}~\bibnamefont {Sarovar}},\ }\href@noop {} {\bibfield  {journal} {\bibinfo
   {journal} {New J. Phys.}\ }\textbf {\bibinfo {volume} {22}},\ \bibinfo
  {pages} {113014} (\bibinfo {year} {2020})}\BibitemShut {NoStop}%
\bibitem [{\citenamefont {Magrini}\ \emph {et~al.}(2021)\citenamefont
  {Magrini}, \citenamefont {Rosenzweig}, \citenamefont {Bach}, \citenamefont
  {Deutschmann-Olek}, \citenamefont {Hofer}, \citenamefont {Hong},
  \citenamefont {Kiesel}, \citenamefont {Kugi},\ and\ \citenamefont
  {Aspelmeyer}}]{magrini2021real}%
  \BibitemOpen
  \bibfield  {author} {\bibinfo {author} {\bibfnamefont {L.}~\bibnamefont
  {Magrini}}, \bibinfo {author} {\bibfnamefont {P.}~\bibnamefont {Rosenzweig}},
  \bibinfo {author} {\bibfnamefont {C.}~\bibnamefont {Bach}}, \bibinfo {author}
  {\bibfnamefont {A.}~\bibnamefont {Deutschmann-Olek}}, \bibinfo {author}
  {\bibfnamefont {S.~G.}\ \bibnamefont {Hofer}}, \bibinfo {author}
  {\bibfnamefont {S.}~\bibnamefont {Hong}}, \bibinfo {author} {\bibfnamefont
  {N.}~\bibnamefont {Kiesel}}, \bibinfo {author} {\bibfnamefont
  {A.}~\bibnamefont {Kugi}}, \ and\ \bibinfo {author} {\bibfnamefont
  {M.}~\bibnamefont {Aspelmeyer}},\ }\href@noop {} {\bibfield  {journal}
  {\bibinfo  {journal} {Nature}\ }\textbf {\bibinfo {volume} {595}},\ \bibinfo
  {pages} {373} (\bibinfo {year} {2021})}\BibitemShut {NoStop}%
\bibitem [{\citenamefont {Motzoi}\ \emph {et~al.}(2016)\citenamefont {Motzoi},
  \citenamefont {Halperin}, \citenamefont {Wang}, \citenamefont {Whaley},\ and\
  \citenamefont {Schirmer}}]{motzoi2016backaction}%
  \BibitemOpen
  \bibfield  {author} {\bibinfo {author} {\bibfnamefont {F.}~\bibnamefont
  {Motzoi}}, \bibinfo {author} {\bibfnamefont {E.}~\bibnamefont {Halperin}},
  \bibinfo {author} {\bibfnamefont {X.}~\bibnamefont {Wang}}, \bibinfo {author}
  {\bibfnamefont {K.~B.}\ \bibnamefont {Whaley}}, \ and\ \bibinfo {author}
  {\bibfnamefont {S.}~\bibnamefont {Schirmer}},\ }\href@noop {} {\bibfield
  {journal} {\bibinfo  {journal} {Phys. Rev. A}\ }\textbf {\bibinfo {volume}
  {94}},\ \bibinfo {pages} {032313} (\bibinfo {year} {2016})}\BibitemShut
  {NoStop}%
\bibitem [{\citenamefont {Basilewitsch}\ \emph {et~al.}(2019)\citenamefont
  {Basilewitsch}, \citenamefont {Cosco}, \citenamefont {Gullo}, \citenamefont
  {M{\"o}tt{\"o}nen}, \citenamefont {Ala-Nissil{\"a}}, \citenamefont {Koch},\
  and\ \citenamefont {Maniscalco}}]{basilewitsch2019reservoir}%
  \BibitemOpen
  \bibfield  {author} {\bibinfo {author} {\bibfnamefont {D.}~\bibnamefont
  {Basilewitsch}}, \bibinfo {author} {\bibfnamefont {F.}~\bibnamefont {Cosco}},
  \bibinfo {author} {\bibfnamefont {N.~L.}\ \bibnamefont {Gullo}}, \bibinfo
  {author} {\bibfnamefont {M.}~\bibnamefont {M{\"o}tt{\"o}nen}}, \bibinfo
  {author} {\bibfnamefont {T.}~\bibnamefont {Ala-Nissil{\"a}}}, \bibinfo
  {author} {\bibfnamefont {C.~P.}\ \bibnamefont {Koch}}, \ and\ \bibinfo
  {author} {\bibfnamefont {S.}~\bibnamefont {Maniscalco}},\ }\href@noop {}
  {\bibfield  {journal} {\bibinfo  {journal} {New J. Phys.}\ }\textbf {\bibinfo
  {volume} {21}},\ \bibinfo {pages} {093054} (\bibinfo {year}
  {2019})}\BibitemShut {NoStop}%
\bibitem [{\citenamefont {Koczor}\ and\ \citenamefont
  {Benjamin}(2020)}]{koczor2020quantum}%
  \BibitemOpen
  \bibfield  {author} {\bibinfo {author} {\bibfnamefont {B.}~\bibnamefont
  {Koczor}}\ and\ \bibinfo {author} {\bibfnamefont {S.~C.}\ \bibnamefont
  {Benjamin}},\ }\href@noop {} {\bibfield  {journal} {\bibinfo  {journal}
  {arXiv preprint arXiv:2008.13774}\ } (\bibinfo {year} {2020})}\BibitemShut
  {NoStop}%
\bibitem [{\citenamefont {Dalgaard}\ \emph
  {et~al.}(2021{\natexlab{b}})\citenamefont {Dalgaard}, \citenamefont
  {Motzoi},\ and\ \citenamefont {Sherson}}]{dalgaard2021predicting}%
  \BibitemOpen
  \bibfield  {author} {\bibinfo {author} {\bibfnamefont {M.}~\bibnamefont
  {Dalgaard}}, \bibinfo {author} {\bibfnamefont {F.}~\bibnamefont {Motzoi}}, \
  and\ \bibinfo {author} {\bibfnamefont {J.}~\bibnamefont {Sherson}},\
  }\href@noop {} {\bibfield  {journal} {\bibinfo  {journal} {arXiv preprint
  arXiv:2107.00008}\ } (\bibinfo {year} {2021}{\natexlab{b}})}\BibitemShut
  {NoStop}%
\bibitem [{\citenamefont {Butcher}(1996)}]{butcher1996history}%
  \BibitemOpen
  \bibfield  {author} {\bibinfo {author} {\bibfnamefont {J.~C.}\ \bibnamefont
  {Butcher}},\ }\href@noop {} {\bibfield  {journal} {\bibinfo  {journal}
  {Applied numerical mathematics}\ }\textbf {\bibinfo {volume} {20}},\ \bibinfo
  {pages} {247} (\bibinfo {year} {1996})}\BibitemShut {NoStop}%
\bibitem [{\citenamefont {S{\o}rensen}\ \emph {et~al.}(2018)\citenamefont
  {S{\o}rensen}, \citenamefont {Aranburu}, \citenamefont {Heinzel},\ and\
  \citenamefont {Sherson}}]{sorensen2018quantum}%
  \BibitemOpen
  \bibfield  {author} {\bibinfo {author} {\bibfnamefont {J.}~\bibnamefont
  {S{\o}rensen}}, \bibinfo {author} {\bibfnamefont {M.}~\bibnamefont
  {Aranburu}}, \bibinfo {author} {\bibfnamefont {T.}~\bibnamefont {Heinzel}}, \
  and\ \bibinfo {author} {\bibfnamefont {J.}~\bibnamefont {Sherson}},\
  }\href@noop {} {\bibfield  {journal} {\bibinfo  {journal} {Phys. Rev. A}\
  }\textbf {\bibinfo {volume} {98}},\ \bibinfo {pages} {022119} (\bibinfo
  {year} {2018})}\BibitemShut {NoStop}%
\bibitem [{\citenamefont {Machnes}\ \emph {et~al.}(2018)\citenamefont
  {Machnes}, \citenamefont {Ass{\'e}mat}, \citenamefont {Tannor},\ and\
  \citenamefont {Wilhelm}}]{machnes2018tunable}%
  \BibitemOpen
  \bibfield  {author} {\bibinfo {author} {\bibfnamefont {S.}~\bibnamefont
  {Machnes}}, \bibinfo {author} {\bibfnamefont {E.}~\bibnamefont
  {Ass{\'e}mat}}, \bibinfo {author} {\bibfnamefont {D.}~\bibnamefont {Tannor}},
  \ and\ \bibinfo {author} {\bibfnamefont {F.~K.}\ \bibnamefont {Wilhelm}},\
  }\href@noop {} {\bibfield  {journal} {\bibinfo  {journal} {Physical review
  letters}\ }\textbf {\bibinfo {volume} {120}},\ \bibinfo {pages} {150401}
  (\bibinfo {year} {2018})}\BibitemShut {NoStop}%
\bibitem [{\citenamefont {Magnus}(1954)}]{magnus1954exponential}%
  \BibitemOpen
  \bibfield  {author} {\bibinfo {author} {\bibfnamefont {W.}~\bibnamefont
  {Magnus}},\ }\href@noop {} {\bibfield  {journal} {\bibinfo  {journal}
  {Communications on pure and applied mathematics}\ }\textbf {\bibinfo {volume}
  {7}},\ \bibinfo {pages} {649} (\bibinfo {year} {1954})}\BibitemShut {NoStop}%
\bibitem [{\citenamefont {Blanes}\ \emph {et~al.}(2009)\citenamefont {Blanes},
  \citenamefont {Casas}, \citenamefont {Oteo},\ and\ \citenamefont
  {Ros}}]{blanes2009magnus}%
  \BibitemOpen
  \bibfield  {author} {\bibinfo {author} {\bibfnamefont {S.}~\bibnamefont
  {Blanes}}, \bibinfo {author} {\bibfnamefont {F.}~\bibnamefont {Casas}},
  \bibinfo {author} {\bibfnamefont {J.-A.}\ \bibnamefont {Oteo}}, \ and\
  \bibinfo {author} {\bibfnamefont {J.}~\bibnamefont {Ros}},\ }\href@noop {}
  {\bibfield  {journal} {\bibinfo  {journal} {Physics reports}\ }\textbf
  {\bibinfo {volume} {470}},\ \bibinfo {pages} {151} (\bibinfo {year}
  {2009})}\BibitemShut {NoStop}%
\bibitem [{\citenamefont {Auer}\ \emph {et~al.}(2018)\citenamefont {Auer},
  \citenamefont {Einkemmer}, \citenamefont {Kandolf},\ and\ \citenamefont
  {Ostermann}}]{auer2018magnus}%
  \BibitemOpen
  \bibfield  {author} {\bibinfo {author} {\bibfnamefont {N.}~\bibnamefont
  {Auer}}, \bibinfo {author} {\bibfnamefont {L.}~\bibnamefont {Einkemmer}},
  \bibinfo {author} {\bibfnamefont {P.}~\bibnamefont {Kandolf}}, \ and\
  \bibinfo {author} {\bibfnamefont {A.}~\bibnamefont {Ostermann}},\ }\href@noop
  {} {\bibfield  {journal} {\bibinfo  {journal} {Computer Physics
  Communications}\ }\textbf {\bibinfo {volume} {228}},\ \bibinfo {pages} {115}
  (\bibinfo {year} {2018})}\BibitemShut {NoStop}%
\bibitem [{\citenamefont {Ma}\ \emph {et~al.}(2020)\citenamefont {Ma},
  \citenamefont {Magann}, \citenamefont {Ho},\ and\ \citenamefont
  {Rabitz}}]{ma2020optimal}%
  \BibitemOpen
  \bibfield  {author} {\bibinfo {author} {\bibfnamefont {A.}~\bibnamefont
  {Ma}}, \bibinfo {author} {\bibfnamefont {A.~B.}\ \bibnamefont {Magann}},
  \bibinfo {author} {\bibfnamefont {T.-S.}\ \bibnamefont {Ho}}, \ and\ \bibinfo
  {author} {\bibfnamefont {H.}~\bibnamefont {Rabitz}},\ }\href@noop {}
  {\bibfield  {journal} {\bibinfo  {journal} {Phys. Rev. A}\ }\textbf {\bibinfo
  {volume} {102}},\ \bibinfo {pages} {013115} (\bibinfo {year}
  {2020})}\BibitemShut {NoStop}%
\bibitem [{\citenamefont {Khaneja}\ \emph {et~al.}(2005)\citenamefont
  {Khaneja}, \citenamefont {Reiss}, \citenamefont {Kehlet}, \citenamefont
  {Schulte-Herbr{\"u}ggen},\ and\ \citenamefont {Glaser}}]{khaneja2005optimal}%
  \BibitemOpen
  \bibfield  {author} {\bibinfo {author} {\bibfnamefont {N.}~\bibnamefont
  {Khaneja}}, \bibinfo {author} {\bibfnamefont {T.}~\bibnamefont {Reiss}},
  \bibinfo {author} {\bibfnamefont {C.}~\bibnamefont {Kehlet}}, \bibinfo
  {author} {\bibfnamefont {T.}~\bibnamefont {Schulte-Herbr{\"u}ggen}}, \ and\
  \bibinfo {author} {\bibfnamefont {S.~J.}\ \bibnamefont {Glaser}},\
  }\href@noop {} {\bibfield  {journal} {\bibinfo  {journal} {Journal of
  magnetic resonance}\ }\textbf {\bibinfo {volume} {172}},\ \bibinfo {pages}
  {296} (\bibinfo {year} {2005})}\BibitemShut {NoStop}%
\bibitem [{\citenamefont {De~Fouquieres}\ \emph {et~al.}(2011)\citenamefont
  {De~Fouquieres}, \citenamefont {Schirmer}, \citenamefont {Glaser},\ and\
  \citenamefont {Kuprov}}]{de2011second}%
  \BibitemOpen
  \bibfield  {author} {\bibinfo {author} {\bibfnamefont {P.}~\bibnamefont
  {De~Fouquieres}}, \bibinfo {author} {\bibfnamefont {S.}~\bibnamefont
  {Schirmer}}, \bibinfo {author} {\bibfnamefont {S.}~\bibnamefont {Glaser}}, \
  and\ \bibinfo {author} {\bibfnamefont {I.}~\bibnamefont {Kuprov}},\
  }\href@noop {} {\bibfield  {journal} {\bibinfo  {journal} {Journal of
  Magnetic Resonance}\ }\textbf {\bibinfo {volume} {212}},\ \bibinfo {pages}
  {412} (\bibinfo {year} {2011})}\BibitemShut {NoStop}%
\bibitem [{\citenamefont {Plischke}\ and\ \citenamefont
  {Bergersen}(1994)}]{plischke1994equilibrium}%
  \BibitemOpen
  \bibfield  {author} {\bibinfo {author} {\bibfnamefont {M.}~\bibnamefont
  {Plischke}}\ and\ \bibinfo {author} {\bibfnamefont {B.}~\bibnamefont
  {Bergersen}},\ }\href@noop {} {\emph {\bibinfo {title} {Equilibrium
  statistical physics}}}\ (\bibinfo  {publisher} {World Scientific},\ \bibinfo
  {year} {1994})\BibitemShut {NoStop}%
\bibitem [{\citenamefont {Cirac}\ and\ \citenamefont
  {Zoller}(1995)}]{cirac1995quantum}%
  \BibitemOpen
  \bibfield  {author} {\bibinfo {author} {\bibfnamefont {J.~I.}\ \bibnamefont
  {Cirac}}\ and\ \bibinfo {author} {\bibfnamefont {P.}~\bibnamefont {Zoller}},\
  }\href@noop {} {\bibfield  {journal} {\bibinfo  {journal} {Phys. Rev. Lett.}\
  }\textbf {\bibinfo {volume} {74}},\ \bibinfo {pages} {4091} (\bibinfo {year}
  {1995})}\BibitemShut {NoStop}%
\bibitem [{\citenamefont {Bloch}(2008)}]{bloch2008quantum}%
  \BibitemOpen
  \bibfield  {author} {\bibinfo {author} {\bibfnamefont {I.}~\bibnamefont
  {Bloch}},\ }\href@noop {} {\bibfield  {journal} {\bibinfo  {journal}
  {Nature}\ }\textbf {\bibinfo {volume} {453}},\ \bibinfo {pages} {1016}
  (\bibinfo {year} {2008})}\BibitemShut {NoStop}%
\bibitem [{\citenamefont {Blais}\ \emph {et~al.}(2020)\citenamefont {Blais},
  \citenamefont {Girvin},\ and\ \citenamefont {Oliver}}]{blais2020quantum}%
  \BibitemOpen
  \bibfield  {author} {\bibinfo {author} {\bibfnamefont {A.}~\bibnamefont
  {Blais}}, \bibinfo {author} {\bibfnamefont {S.~M.}\ \bibnamefont {Girvin}}, \
  and\ \bibinfo {author} {\bibfnamefont {W.~D.}\ \bibnamefont {Oliver}},\
  }\href@noop {} {\bibfield  {journal} {\bibinfo  {journal} {Nature Physics}\
  }\textbf {\bibinfo {volume} {16}},\ \bibinfo {pages} {247} (\bibinfo {year}
  {2020})}\BibitemShut {NoStop}%
\bibitem [{\citenamefont {Wu}\ \emph {et~al.}(2020)\citenamefont {Wu},
  \citenamefont {Liang}, \citenamefont {Tian}, \citenamefont {Yang},
  \citenamefont {Chen}, \citenamefont {Liu}, \citenamefont {Tey},\ and\
  \citenamefont {You}}]{wu2020concise}%
  \BibitemOpen
  \bibfield  {author} {\bibinfo {author} {\bibfnamefont {X.}~\bibnamefont
  {Wu}}, \bibinfo {author} {\bibfnamefont {X.}~\bibnamefont {Liang}}, \bibinfo
  {author} {\bibfnamefont {Y.}~\bibnamefont {Tian}}, \bibinfo {author}
  {\bibfnamefont {F.}~\bibnamefont {Yang}}, \bibinfo {author} {\bibfnamefont
  {C.}~\bibnamefont {Chen}}, \bibinfo {author} {\bibfnamefont {Y.-C.}\
  \bibnamefont {Liu}}, \bibinfo {author} {\bibfnamefont {M.~K.}\ \bibnamefont
  {Tey}}, \ and\ \bibinfo {author} {\bibfnamefont {L.}~\bibnamefont {You}},\
  }\href@noop {} {\bibfield  {journal} {\bibinfo  {journal} {Chinese Physics
  B}\ } (\bibinfo {year} {2020})}\BibitemShut {NoStop}%
\bibitem [{\citenamefont {Kinos}\ \emph {et~al.}(2021)\citenamefont {Kinos},
  \citenamefont {Hunger}, \citenamefont {Kolesov}, \citenamefont {M{\o}lmer},
  \citenamefont {de~Riedmatten}, \citenamefont {Goldner}, \citenamefont
  {Tallaire}, \citenamefont {Morvan}, \citenamefont {Berger}, \citenamefont
  {Welinski} \emph {et~al.}}]{kinos2021roadmap}%
  \BibitemOpen
  \bibfield  {author} {\bibinfo {author} {\bibfnamefont {A.}~\bibnamefont
  {Kinos}}, \bibinfo {author} {\bibfnamefont {D.}~\bibnamefont {Hunger}},
  \bibinfo {author} {\bibfnamefont {R.}~\bibnamefont {Kolesov}}, \bibinfo
  {author} {\bibfnamefont {K.}~\bibnamefont {M{\o}lmer}}, \bibinfo {author}
  {\bibfnamefont {H.}~\bibnamefont {de~Riedmatten}}, \bibinfo {author}
  {\bibfnamefont {P.}~\bibnamefont {Goldner}}, \bibinfo {author} {\bibfnamefont
  {A.}~\bibnamefont {Tallaire}}, \bibinfo {author} {\bibfnamefont
  {L.}~\bibnamefont {Morvan}}, \bibinfo {author} {\bibfnamefont
  {P.}~\bibnamefont {Berger}}, \bibinfo {author} {\bibfnamefont
  {S.}~\bibnamefont {Welinski}},  \emph {et~al.},\ }\href@noop {} {\bibfield
  {journal} {\bibinfo  {journal} {arXiv preprint arXiv:2103.15743}\ } (\bibinfo
  {year} {2021})}\BibitemShut {NoStop}%
\bibitem [{\citenamefont {Fowler}\ \emph {et~al.}(2012)\citenamefont {Fowler},
  \citenamefont {Mariantoni}, \citenamefont {Martinis},\ and\ \citenamefont
  {Cleland}}]{fowler2012surface}%
  \BibitemOpen
  \bibfield  {author} {\bibinfo {author} {\bibfnamefont {A.~G.}\ \bibnamefont
  {Fowler}}, \bibinfo {author} {\bibfnamefont {M.}~\bibnamefont {Mariantoni}},
  \bibinfo {author} {\bibfnamefont {J.~M.}\ \bibnamefont {Martinis}}, \ and\
  \bibinfo {author} {\bibfnamefont {A.~N.}\ \bibnamefont {Cleland}},\
  }\href@noop {} {\bibfield  {journal} {\bibinfo  {journal} {Phys. Rev. A}\
  }\textbf {\bibinfo {volume} {86}},\ \bibinfo {pages} {032324} (\bibinfo
  {year} {2012})}\BibitemShut {NoStop}%
\bibitem [{\citenamefont {Barends}\ \emph {et~al.}(2014)\citenamefont
  {Barends}, \citenamefont {Kelly}, \citenamefont {Megrant}, \citenamefont
  {Veitia}, \citenamefont {Sank}, \citenamefont {Jeffrey}, \citenamefont
  {White}, \citenamefont {Mutus}, \citenamefont {Fowler}, \citenamefont
  {Campbell} \emph {et~al.}}]{barends2014superconducting}%
  \BibitemOpen
  \bibfield  {author} {\bibinfo {author} {\bibfnamefont {R.}~\bibnamefont
  {Barends}}, \bibinfo {author} {\bibfnamefont {J.}~\bibnamefont {Kelly}},
  \bibinfo {author} {\bibfnamefont {A.}~\bibnamefont {Megrant}}, \bibinfo
  {author} {\bibfnamefont {A.}~\bibnamefont {Veitia}}, \bibinfo {author}
  {\bibfnamefont {D.}~\bibnamefont {Sank}}, \bibinfo {author} {\bibfnamefont
  {E.}~\bibnamefont {Jeffrey}}, \bibinfo {author} {\bibfnamefont {T.~C.}\
  \bibnamefont {White}}, \bibinfo {author} {\bibfnamefont {J.}~\bibnamefont
  {Mutus}}, \bibinfo {author} {\bibfnamefont {A.~G.}\ \bibnamefont {Fowler}},
  \bibinfo {author} {\bibfnamefont {B.}~\bibnamefont {Campbell}},  \emph
  {et~al.},\ }\href@noop {} {\bibfield  {journal} {\bibinfo  {journal}
  {Nature}\ }\textbf {\bibinfo {volume} {508}},\ \bibinfo {pages} {500}
  (\bibinfo {year} {2014})}\BibitemShut {NoStop}%
\bibitem [{\citenamefont {Christandl}\ \emph {et~al.}(2004)\citenamefont
  {Christandl}, \citenamefont {Datta}, \citenamefont {Ekert},\ and\
  \citenamefont {Landahl}}]{ChristandlStateTransfer}%
  \BibitemOpen
  \bibfield  {author} {\bibinfo {author} {\bibfnamefont {M.}~\bibnamefont
  {Christandl}}, \bibinfo {author} {\bibfnamefont {N.}~\bibnamefont {Datta}},
  \bibinfo {author} {\bibfnamefont {A.}~\bibnamefont {Ekert}}, \ and\ \bibinfo
  {author} {\bibfnamefont {A.~J.}\ \bibnamefont {Landahl}},\ }\href {\doibase
  10.1103/PhysRevLett.92.187902} {\bibfield  {journal} {\bibinfo  {journal}
  {Phys. Rev. Lett.}\ }\textbf {\bibinfo {volume} {92}},\ \bibinfo {pages}
  {187902} (\bibinfo {year} {2004})}\BibitemShut {NoStop}%
\bibitem [{\citenamefont {Yung}(2006)}]{yung2006quantum}%
  \BibitemOpen
  \bibfield  {author} {\bibinfo {author} {\bibfnamefont {M.-H.}\ \bibnamefont
  {Yung}},\ }\href@noop {} {\bibfield  {journal} {\bibinfo  {journal} {Phys.
  Rev. A}\ }\textbf {\bibinfo {volume} {74}},\ \bibinfo {pages} {030303}
  (\bibinfo {year} {2006})}\BibitemShut {NoStop}%
\bibitem [{\citenamefont {Machnes}\ \emph {et~al.}(2011)\citenamefont
  {Machnes}, \citenamefont {Sander}, \citenamefont {Glaser}, \citenamefont
  {De~Fouqui{\`e}res}, \citenamefont {Gruslys}, \citenamefont {Schirmer},\ and\
  \citenamefont {Schulte-Herbr{\"u}ggen}}]{machnes2011comparing}%
  \BibitemOpen
  \bibfield  {author} {\bibinfo {author} {\bibfnamefont {S.}~\bibnamefont
  {Machnes}}, \bibinfo {author} {\bibfnamefont {U.}~\bibnamefont {Sander}},
  \bibinfo {author} {\bibfnamefont {S.~J.}\ \bibnamefont {Glaser}}, \bibinfo
  {author} {\bibfnamefont {P.}~\bibnamefont {De~Fouqui{\`e}res}}, \bibinfo
  {author} {\bibfnamefont {A.}~\bibnamefont {Gruslys}}, \bibinfo {author}
  {\bibfnamefont {S.}~\bibnamefont {Schirmer}}, \ and\ \bibinfo {author}
  {\bibfnamefont {T.}~\bibnamefont {Schulte-Herbr{\"u}ggen}},\ }\href@noop {}
  {\bibfield  {journal} {\bibinfo  {journal} {Phys. Rev. A}\ }\textbf {\bibinfo
  {volume} {84}},\ \bibinfo {pages} {022305} (\bibinfo {year}
  {2011})}\BibitemShut {NoStop}%
\bibitem [{\citenamefont {Motzoi}\ \emph {et~al.}(2011)\citenamefont {Motzoi},
  \citenamefont {Gambetta}, \citenamefont {Merkel},\ and\ \citenamefont
  {Wilhelm}}]{motzoi2011optimal}%
  \BibitemOpen
  \bibfield  {author} {\bibinfo {author} {\bibfnamefont {F.}~\bibnamefont
  {Motzoi}}, \bibinfo {author} {\bibfnamefont {J.~M.}\ \bibnamefont
  {Gambetta}}, \bibinfo {author} {\bibfnamefont {S.}~\bibnamefont {Merkel}}, \
  and\ \bibinfo {author} {\bibfnamefont {F.}~\bibnamefont {Wilhelm}},\
  }\href@noop {} {\bibfield  {journal} {\bibinfo  {journal} {Phys. Rev. A}\
  }\textbf {\bibinfo {volume} {84}},\ \bibinfo {pages} {022307} (\bibinfo
  {year} {2011})}\BibitemShut {NoStop}%
\bibitem [{\citenamefont {Schulte-Herbr{\"u}ggen}\ \emph
  {et~al.}(2011)\citenamefont {Schulte-Herbr{\"u}ggen}, \citenamefont
  {Sp{\"o}rl}, \citenamefont {Khaneja},\ and\ \citenamefont
  {Glaser}}]{schulte2011optimal}%
  \BibitemOpen
  \bibfield  {author} {\bibinfo {author} {\bibfnamefont {T.}~\bibnamefont
  {Schulte-Herbr{\"u}ggen}}, \bibinfo {author} {\bibfnamefont {A.}~\bibnamefont
  {Sp{\"o}rl}}, \bibinfo {author} {\bibfnamefont {N.}~\bibnamefont {Khaneja}},
  \ and\ \bibinfo {author} {\bibfnamefont {S.}~\bibnamefont {Glaser}},\
  }\href@noop {} {\bibfield  {journal} {\bibinfo  {journal} {Journal of Physics
  B: Atomic, Molecular and Optical Physics}\ }\textbf {\bibinfo {volume}
  {44}},\ \bibinfo {pages} {154013} (\bibinfo {year} {2011})}\BibitemShut
  {NoStop}%
\bibitem [{\citenamefont {Lucarelli}(2018)}]{Lucarelli2018}%
  \BibitemOpen
  \bibfield  {author} {\bibinfo {author} {\bibfnamefont {D.}~\bibnamefont
  {Lucarelli}},\ }\href@noop {} {\bibfield  {journal} {\bibinfo  {journal}
  {Phys. Rev. A}\ }\textbf {\bibinfo {volume} {97}},\ \bibinfo {pages} {062346}
  (\bibinfo {year} {2018})}\BibitemShut {NoStop}%
\bibitem [{\citenamefont {Caneva}\ \emph {et~al.}(2011)\citenamefont {Caneva},
  \citenamefont {Calarco},\ and\ \citenamefont
  {Montangero}}]{caneva2011chopped}%
  \BibitemOpen
  \bibfield  {author} {\bibinfo {author} {\bibfnamefont {T.}~\bibnamefont
  {Caneva}}, \bibinfo {author} {\bibfnamefont {T.}~\bibnamefont {Calarco}}, \
  and\ \bibinfo {author} {\bibfnamefont {S.}~\bibnamefont {Montangero}},\
  }\href@noop {} {\bibfield  {journal} {\bibinfo  {journal} {Phys. Rev. A}\
  }\textbf {\bibinfo {volume} {84}},\ \bibinfo {pages} {022326} (\bibinfo
  {year} {2011})}\BibitemShut {NoStop}%
\bibitem [{\citenamefont {van Frank}\ \emph {et~al.}(2016)\citenamefont {van
  Frank}, \citenamefont {Bonneau}, \citenamefont {Schmiedmayer}, \citenamefont
  {Hild}, \citenamefont {Gross}, \citenamefont {Cheneau}, \citenamefont
  {Bloch}, \citenamefont {Pichler}, \citenamefont {Negretti}, \citenamefont
  {Calarco} \emph {et~al.}}]{van2016optimal}%
  \BibitemOpen
  \bibfield  {author} {\bibinfo {author} {\bibfnamefont {S.}~\bibnamefont {van
  Frank}}, \bibinfo {author} {\bibfnamefont {M.}~\bibnamefont {Bonneau}},
  \bibinfo {author} {\bibfnamefont {J.}~\bibnamefont {Schmiedmayer}}, \bibinfo
  {author} {\bibfnamefont {S.}~\bibnamefont {Hild}}, \bibinfo {author}
  {\bibfnamefont {C.}~\bibnamefont {Gross}}, \bibinfo {author} {\bibfnamefont
  {M.}~\bibnamefont {Cheneau}}, \bibinfo {author} {\bibfnamefont
  {I.}~\bibnamefont {Bloch}}, \bibinfo {author} {\bibfnamefont
  {T.}~\bibnamefont {Pichler}}, \bibinfo {author} {\bibfnamefont
  {A.}~\bibnamefont {Negretti}}, \bibinfo {author} {\bibfnamefont
  {T.}~\bibnamefont {Calarco}},  \emph {et~al.},\ }\href@noop {} {\bibfield
  {journal} {\bibinfo  {journal} {Scientific reports}\ }\textbf {\bibinfo
  {volume} {6}},\ \bibinfo {pages} {34187} (\bibinfo {year}
  {2016})}\BibitemShut {NoStop}%
\bibitem [{\citenamefont {Kraft}(1988)}]{kraft1988software}%
  \BibitemOpen
  \bibfield  {author} {\bibinfo {author} {\bibfnamefont {D.}~\bibnamefont
  {Kraft}},\ }\href@noop {} {\bibfield  {journal} {\bibinfo  {journal} {DFVLR
  Obersfaffeuhofen, Germany}\ } (\bibinfo {year} {1988})}\BibitemShut {NoStop}%
\bibitem [{\citenamefont {Virtanen}\ \emph {et~al.}(2020)\citenamefont
  {Virtanen}, \citenamefont {Gommers}, \citenamefont {Oliphant}, \citenamefont
  {Haberland}, \citenamefont {Reddy}, \citenamefont {Cournapeau}, \citenamefont
  {Burovski}, \citenamefont {Peterson}, \citenamefont {Weckesser},
  \citenamefont {Bright} \emph {et~al.}}]{virtanen2020scipy}%
  \BibitemOpen
  \bibfield  {author} {\bibinfo {author} {\bibfnamefont {P.}~\bibnamefont
  {Virtanen}}, \bibinfo {author} {\bibfnamefont {R.}~\bibnamefont {Gommers}},
  \bibinfo {author} {\bibfnamefont {T.~E.}\ \bibnamefont {Oliphant}}, \bibinfo
  {author} {\bibfnamefont {M.}~\bibnamefont {Haberland}}, \bibinfo {author}
  {\bibfnamefont {T.}~\bibnamefont {Reddy}}, \bibinfo {author} {\bibfnamefont
  {D.}~\bibnamefont {Cournapeau}}, \bibinfo {author} {\bibfnamefont
  {E.}~\bibnamefont {Burovski}}, \bibinfo {author} {\bibfnamefont
  {P.}~\bibnamefont {Peterson}}, \bibinfo {author} {\bibfnamefont
  {W.}~\bibnamefont {Weckesser}}, \bibinfo {author} {\bibfnamefont
  {J.}~\bibnamefont {Bright}},  \emph {et~al.},\ }\href@noop {} {\bibfield
  {journal} {\bibinfo  {journal} {Nature methods}\ }\textbf {\bibinfo {volume}
  {17}},\ \bibinfo {pages} {261} (\bibinfo {year} {2020})}\BibitemShut
  {NoStop}%
\bibitem [{\citenamefont {Arnal}\ \emph {et~al.}(2018)\citenamefont {Arnal},
  \citenamefont {Casas},\ and\ \citenamefont {Chiralt}}]{arnal2018general}%
  \BibitemOpen
  \bibfield  {author} {\bibinfo {author} {\bibfnamefont {A.}~\bibnamefont
  {Arnal}}, \bibinfo {author} {\bibfnamefont {F.}~\bibnamefont {Casas}}, \ and\
  \bibinfo {author} {\bibfnamefont {C.}~\bibnamefont {Chiralt}},\ }\href@noop
  {} {\bibfield  {journal} {\bibinfo  {journal} {Journal of Physics
  Communications}\ }\textbf {\bibinfo {volume} {2}},\ \bibinfo {pages} {035024}
  (\bibinfo {year} {2018})}\BibitemShut {NoStop}%
\end{thebibliography}%

\end{document}